\documentclass[aps,prb,twocolumn,amsmath,amssymb]{revtex4}
\usepackage{graphicx}
\usepackage{bm}
\usepackage{bbold}
\usepackage{hhline}

\def\br{ \bm{r} }
\def\bk{ \bm{k} }
\def\bmd{ \bm{d} }
\def\be{ \bm{e} }
\def\spup{ \lvert\uparrow\rangle }
\def\spdown{ \lvert\downarrow\rangle }
\def\hbx{ \hat{\bm{x}} }
\def\hby{ \hat{\bm{y}} }
\def\hbz{ \hat{\bm{z}} }
\def\hbm{ \hat{\bm{m}} }
\def\bB{ \bm{B} }
\def\bH{ \bm{H} }
\def\bmu{ \bm{\mu} }
\def\bmb{ \bm{b} }

\def\re{ \,\mathrm{Re}\, }
\def\sign{ \mathrm{sign} }
\def\tr{ \,\mathrm{tr}\, }

\begin{document}
\title{Spin susceptibility of superconductors with strong spin-orbit coupling}

\author{K. V. Samokhin\footnote{E-mail: kirill.samokhin@brocku.ca}}
\affiliation{Department of Physics, Brock University, St. Catharines, Ontario L2S 3A1, Canada}


\begin{abstract}
We show that in some trigonal and hexagonal crystals the Zeeman coupling of band electrons with an external magnetic field is strongly anisotropic and necessarily vanishes along the main symmetry axis. 
This leads to qualitative changes in the temperature dependence of the electron spin susceptibility in the superconducting state. In particular, the power-law exponents at low temperatures due
to the contribution of the nodal quasiparticles are considerably modified compared to their textbook values.
\end{abstract}

\maketitle

\section{Introduction}
\label{sec: Intro}

Measuring the spin magnetic response is an important characterization tool of superconductors, both conventional and unconventional.\cite{SU-review,TheBook} 
The usual assumption is that the Zeeman interaction of the conduction electrons with an external magnetic field has essentially the same form as for bare electrons, i.e., is independent of the wave vector. 
Then, the temperature behaviour of the spin susceptibility depends only on the superconducting gap structure, but not on the character of the electron Bloch bands. 
If the superconducting pairing occurs in the spin-singlet channel, then the susceptibility tensor 
$\chi_{ij}(T)=\chi(T)\delta_{ij}$ is entirely determined by the thermally excited quasiparticles and
in a fully gapped superconducting state one would see an exponentially decreasing $\chi(T)$ at low temperatures. In contrast, if the gap has zeros on the Fermi surface, then $\chi(T)\propto T^2$ for isolated first-order point
nodes, or $\chi(T)\propto T$ for line nodes or second-order point nodes. In the triplet case, the susceptibility acquires a nontrivial tensor structure depending on the mutual orientation of the order parameter $\bm{d}$ 
and the external magnetic field $\bH$. For $\bm{d}\parallel\bH$ the spin response is determined only by the excitations and the susceptibility has the same temperature dependence as in the singlet case, 
whereas for $\bm{d}\perp\bH$ both the Cooper pairs and the excitations contribute to the spin response and the susceptibility is unchanged compared to the normal state.

The electron-lattice spin-orbit (SO) coupling in many superconductors of current interest, in particular, the heavy-fermion compounds, is strong compared with the Zeeman energy and the energy scales associated with 
superconductivity, and one can call into question the validity of the assumption 
that the effective Zeeman interaction in the Bloch bands has the same simple form $\hat H_Z=\mu_B\bH\hat{\bm\sigma}$ as for bare electrons (here the spin magnetic moment is equal to $-\mu_B\hat{\bm\sigma}$, 
$\mu_B$ is the Bohr magneton,
the electron charge is $-e$, the Land\'e factor is set to $2$, and $\hat{\bm\sigma}$ are the Pauli matrices). Even in the presence of the SO coupling,
the Bloch bands in a centrosymmetric crystal are twofold degenerate.\cite{Kittel-book} Focusing on just one band, one can write the effective Zeeman Hamiltonian at the wave vector 
$\bk$ in the general form as $\hat H_Z=-\hat{\bm m}(\bk)\bH$, where the Hermitian $2\times 2$ matrix $\hat{\bm m}$ has the meaning of the magnetic moment of the band electrons.\cite{CB60,Roth66,LL-9} 

While the effects of the Zeeman coupling anisotropy due to the SO coupling have been extensively studied in the context of semiconductors, see, e.g., Ref. \onlinecite{Winkler-book}, 
its consequences for the spin response of superconductors 
have received relatively little attention in the literature. These consequences are expected to be profound, especially in the light of the recent developments in the field,\cite{WW09,j32,Fis13,RS16,NHI16,Sam19,Sam19-PRB} 
which demonstrated that the SO coupling can significantly affect the symmetry of the Bloch bands and invalidate the pseudospin picture\cite{And84,UR85} commonly used in the theory of unconventional superconductivity.

The pseudospin approach is based on the assumption that the twofold degenerate Bloch states transform under the crystal point group operations in the same way as the pure spin states $\spup$ and $\spdown$ at each wave vector, 
including the center of the Brillouin zone (the $\Gamma$ point). However, the latter cannot always be true, because, mathematically, not all double-valued irreducible representations of the crystal point group 
(or, more accurately, corepresentations of the magnetic point group) are equivalent to the spin-1/2 representation. We will show how a ``non-pseudospin'' character of the Bloch states leads to a strongly 
anisotropic Zeeman interaction in some bands in trigonal and hexagonal crystals.

Our main goal is to calculate the spin susceptibility of superconductors with an anisotropic Zeeman coupling, whose form is constrained only by the symmetry of the system. Impurities are neglected, as are the effects 
of the interaction of the electron charges with the magnetic field. In Secs. \ref{sec: Zeeman coupling} and \ref{sec: intraband Zeeman}, we develop the general symmetry-based theory of the spin response of the Bloch bands 
in crystals with the SO coupling and identify the cases in which the Zeeman interaction is qualitatively different from the usual isotropic expression. In Sec. \ref{sec: SC state}, the spin response in the 
superconducting state is studied and the spin susceptibility of a general multiband superconductor is derived. In Sec. \ref{sec: hexagonal}, we discuss applications to singlet and triplet
superconductors of hexagonal symmetry. Some technical details are relegated to the Appendices.  
Throughout the paper we use the units in which $\hbar=1$ and $k_B=1$, neglecting, in particular, the difference between the quasiparticle momentum and the wave vector.

\section{Zeeman coupling in the band representation}
\label{sec: Zeeman coupling}

We consider a centrosymmetric nonmagnetic crystal with $N$ twofold degenerate bands crossing the Fermi level. The Bloch states are degenerate at each wave vector 
$\bk$ due to the conjugation symmetry ${\cal C}=KI$ (Ref. \onlinecite{Kittel-book}): the states $|\bk,n,1\rangle$ and $|\bk,n,2\rangle\equiv{\cal C}|\bk,n,1\rangle$ are orthogonal and have the same energy. 
Here $K$ is the time reversal (TR) operation, $I$ is the space inversion, $n=1,...,N$ labels the bands, and the index $s=1,2$, which distinguishes two degenerate states within the same band, is called the 
conjugation index. The conjugation operation is antiunitary and, since the Bloch states are spin-$1/2$ spinors, we have ${\cal C}|\bk,n,2\rangle=-|\bk,n,1\rangle$, i.e., ${\cal C}^2=-1$. 
The conjugation degeneracy, which is also called the Kramers degeneracy, can be lifted by an external magnetic field. Note that the assumption of inversion symmetry is important: in a non-centrosymmetric crystal 
the band degeneracy is lifted by the SO coupling even at zero magnetic field, with wide-reaching implications for superconductivity.\cite{NCSC-book} 

The Zeeman coupling of electrons with the magnetic field $\bm{H}$ is described by the Hamiltonian $\hat H_Z=-\bH\hbm$, where $(\hat m_x,\hat m_y,\hat m_z)=-\mu_B(\hat\sigma_1,\hat\sigma_2,\hat\sigma_3)$ 
is the operator of the spin magnetic moment. 
In the band representation, the Zeeman Hamiltonian is diagonal in the wave vector, but not in the band and conjugation indices:
\begin{equation}
\label{H_Z-matrix element}
  \langle\bk,n,s|\hat H_Z|\bk,n',s'\rangle=-\bH\bm{m}_{nn',ss'}(\bk).
\end{equation}
The matrix elements of the spin magnetic moment in the Bloch basis can be represented as linear combinations of the unit matrix and the Pauli matrices in the conjugacy space as follows:
\begin{eqnarray}
\label{m-matrix elements}
  m_{nn',ss',j}(\bk) &\equiv& \langle\bk,n,s|\hat m_j|\bk,n',s'\rangle \nonumber\\
  &=& iA_{nn',j}(\bk)\delta_{ss'}+\bB_{nn',j}(\bk)\bm{\sigma}_{ss'}.
\end{eqnarray}
The intraband and interband components correspond to $n=n'$ and $n\neq n'$, respectively. 

In principle, the functions $A$ and $\bB$ in Eq. (\ref{m-matrix elements}) can be calculated if a treatable model of the band structure is available. Here we adopt a different approach: without resorting to any particular 
microscopic model, we regard $A$ and $\bB$ as phenomenological parameters, whose form is only constrained by the symmetries of the system. In particular, from the requirement that $\hbm$ is a Hermitian operator, 
we have $\langle\psi|\hbm|\psi'\rangle=\langle\psi'|\hbm|\psi\rangle^*$, therefore,
\begin{equation}
\label{AB-constraint-Hermite}
  A_{nn',i}(\bk)=-A_{n'n,i}^*(\bk),\quad \bB_{nn',i}(\bk)=\bB_{n'n,i}^*(\bk).
\end{equation}
Additional constraints are obtained by analyzing the response of the spin magnetic moment to TR and the point group operations, which in turn depends on the transformation properties of the Bloch states.

From the group-theoretical point of view, the spinor Bloch states $|\bk,n,1\rangle$ and $|\bk,n,2\rangle$ form the basis of an irreducible double-valued two-dimensional (2D) corepresentation (corep) of the magnetic point group 
of the wave vector $\bk$, see Appendix \ref{app: Bloch bases}. Four-dimensional coreps, which are possible in crystals of cubic symmetry, are not considered here. 
If the corep at the $\Gamma$ point is described by $2\times 2$ matrices $\hat{\cal D}_n(g)$, where $g$ is an element of the crystal point group $\mathbb{G}$, 
then one can construct the Bloch bases at $\bk\neq\bm{0}$ using the following expression:\cite{Sam19,Sam19-PRB}
\begin{equation}
\label{general-prescription}
    g|\bk,n,s\rangle=\sum_{s'}|g\bk,n,s'\rangle {\cal D}_{n,s's}(g),
\end{equation}
which defines the basis at the wave vector $g\bk$ given the basis at $\bk$. 

In a ``pseudospin'' band, the $\Gamma$-point corep is equivalent to the spin-$1/2$ corep and the conjugate Bloch states 
affected by the SO coupling respond to the point group operations in exactly the same way as the pure spin states. Therefore, one can put $\hat{\cal D}_n(g)=\hat D^{(1/2)}(R)$ for both proper rotations ($g=R$) 
and improper rotations ($g=IR$). Here 
\begin{equation}
\label{D-spin}
  \hat D^{(1/2)}(R)=e^{-i\theta(\bm{n}\hat{\bm{\sigma}})/2}
\end{equation}
is the spin-1/2 representation of a counterclockwise rotation $R$ through an angle $\theta$ about an axis $\bm{n}$. Thus, in a pseudospin band Eq. (\ref{general-prescription}) reproduces the well-known Ueda-Rice formula.\cite{UR85} 

In general, however, the Bloch states at the $\Gamma$ point correspond to a corep which is not equivalent to the spin-$1/2$ corep. Therefore, $\hat{\cal D}_n(g)\neq\hat D^{(1/2)}(R)$ for some $g$ and the Ueda-Rice prescription 
does not work. In this case, the band is called a ``non-pseudospin'' band. The corep matrices in twofold degenerate non-pseudospin bands are shown in Table \ref{table: non-pseudospin-corep-matrices}, 
see Appendix \ref{app: Bloch bases} for details. 

\begin{table*}
\caption{The $\Gamma$-point corepresentation matrices in twofold degenerate non-pseudospin bands; $R$ denotes the rotational generators of the point group $\mathbb{G}$ and the spin rotation matrix $\hat D^{(1/2)}(R)$ is
given by Eq. (\ref{D-spin}). The notations for the point group elements are the same as in Ref. \onlinecite{LL-3}.}
{\begin{tabular}{lll}
    \toprule
    $\mathbb{G}$ &  corep & $\hat{\cal D}(R)$ \\ \colrule
    $\mathbf{C}_{4h}$  & $(\Gamma_7,\Gamma_8)$  &  $\hat{\cal D}(C_{4z})=-\hat D^{(1/2)}(C_{4z})$ \\ \colrule
    $\mathbf{D}_{4h}$ & $\Gamma_7$  & $\hat{\cal D}(C_{4z})=-\hat D^{(1/2)}(C_{4z})$,\quad $\hat{\cal D}(C_{2y})=\hat D^{(1/2)}(C_{2y})$ \\ \colrule
    $\mathbf{C}_{3i}$  & $\Gamma_6$ &  $\hat{\cal D}(C_{3z})=-\hat\sigma_0$ \\ \colrule
    $\mathbf{D}_{3d}$  & $(\Gamma_5,\Gamma_6)$  & $\hat{\cal D}(C_{3z})=-\hat\sigma_0$,\quad $\hat{\cal D}(C_{2y})=\hat D^{(1/2)}(C_{2y})$ \\ \colrule								
    $\mathbf{C}_{6h}$  & $(\Gamma_{9},\Gamma_{10})$  & $\hat{\cal D}(C_{6z})=-\hat D^{(1/2)}(C_{6z})$ \\ 
                      &  $(\Gamma_{11},\Gamma_{12})$\hspace*{3mm}  & $\hat{\cal D}(C_{6z})=-\hat D^{(1/2)}(C_{2z})$ \\ \colrule
    $\mathbf{D}_{6h}$\hspace*{2mm}  & $\Gamma_8$  & $\hat{\cal D}(C_{6z})=-\hat D^{(1/2)}(C_{6z})$,\quad $\hat{\cal D}(C_{2y})=\hat D^{(1/2)}(C_{2y})$ \\ 
                       & $\Gamma_9$ & $\hat{\cal D}(C_{6z})=-\hat D^{(1/2)}(C_{2z})$,\quad $\hat{\cal D}(C_{2y})=\hat D^{(1/2)}(C_{2y})$ \\ \colrule                                                             
    $\mathbf{O}_h$  & $\Gamma_7$ &  $\hat{\cal D}(C_{4z})=-\hat D^{(1/2)}(C_{4z})$,\quad $\hat{\cal D}(C_{2y})=\hat D^{(1/2)}(C_{2y})$,\quad $\hat{\cal D}(C_{3xyz})=\hat D^{(1/2)}(C_{3xyz})$  \\ \botrule                                                           
\end{tabular}}
\label{table: non-pseudospin-corep-matrices}
\end{table*}

It follows from Eq. (\ref{general-prescription}) that under the space inversion the conjugate Bloch states transform as $I|\bk,n,s\rangle=p_n|-\bk,n,s\rangle$,
where $p_n=\pm$ denotes the parity of the band. The pseudovector quantity $\hbm$ must remain invariant under inversion, $I\hbm I^{-1}=\hbm$, therefore
$$
  \langle\bk,n,s|\hbm|\bk,n',s'\rangle=p_np_{n'}\langle-\bk,n,s|\hbm|-\bk,n',s'\rangle.
$$
Using the representation (\ref{m-matrix elements}), we find that the parity of $A$ and $\bB$ is determined by the relative parity of the bands $p_np_{n'}$. For the transformation under the TR operation $K={\cal C}I$, we have
\begin{equation}
\label{general-prescription-K}
    \begin{array}{l}
    K|\bk,n,1\rangle=p_n|-\bk,n,2\rangle,\medskip\\ 
    K|\bk,n,2\rangle=-p_n|-\bk,n,1\rangle.
    \end{array}
\end{equation}
Being a magnetic moment operator, $\hbm$ must change sign under TR: $K\hbm K^{-1}=-\hbm$, therefore
\begin{eqnarray*}
  && \langle\bk,n,s|\hbm|\bk,n',s'\rangle=-\langle\bk,n,s|K^\dagger\hbm K|\bk,n',s'\rangle \\
  && =-p_np_{n'}\sum_{s_1s_2}\sigma_{2,s's_1}\langle-\bk,n',s_1|\hbm|-\bk,n,s_2\rangle\sigma_{2,s_2s}.
\end{eqnarray*}
In the second line here we used Eq. (\ref{general-prescription-K}) and the expression $\langle\psi|K^\dagger|\psi'\rangle=\langle\psi'|K|\psi\rangle$, which follows from the antiunitarity of $K$.
In terms of $A$ and $\bB$, we obtain: $A_{nn',i}(\bk)=-p_np_{n'}A_{n'n,i}(-\bk)$ and $\bB_{nn',i}(\bk)=p_np_{n'}\bB_{n'n,i}(-\bk)$.
Putting all together, we find that the functions $A$ and $\bB$ are real and satisfy
\begin{equation}
\label{A-constraints}
  \begin{array}{c}
  A_{nn',i}(\bk)=-A_{n'n,i}(\bk), \medskip\\ 
  A_{nn',i}(-\bk)=p_np_{n'}A_{nn',i}(\bk),
  \end{array}
\end{equation}
and
\begin{equation}
\label{B-constraints}
  \begin{array}{c}
  \bB_{nn',i}(\bk)=\bB_{n'n,i}(\bk),\medskip\\ 
  \bB_{nn',i}(-\bk)=p_np_{n'}\bB_{nn',i}(\bk).
  \end{array}
\end{equation}
Additional constraints are imposed by the requirement of invariance under the remaining elements of the crystal point group. 

Under a point-group operation $g$ (e.g., a rotation), an arbitrary state of the system is transformed into another state as $|\psi\rangle\to|\tilde\psi\rangle=g|\psi\rangle$ and the magnetic field is transformed into 
$g\bH$. The matrix elements of the Zeeman Hamiltonian $\hat H_Z(\bH)$ in the basis $|\psi\rangle$ are equal to the matrix elements of the Hamiltonian $\hat H_Z(g\bH)$ in the transformed basis $|\tilde\psi\rangle$.
Therefore,
$$
  \langle\bk,n,s|(\hbm\cdot\bH)|\bk,n',s'\rangle=\langle\bk,n,s|g^\dagger(\hbm\cdot g\bH)g|\bk,n',s'\rangle,
$$
see, e.g., Ref. \onlinecite{Bir-Pikus-book}. Using Eqs. (\ref{m-matrix elements}) and (\ref{general-prescription}), the invariance condition under $g\in\mathbb{G}$ takes the following form:
\begin{eqnarray}
\label{AB-invariance-g}
  && iA_{nn',j}(\bk)\hat\sigma_0+\bB_{nn',j}(\bk)\hat{\bm{\sigma}} \nonumber\\
  && \quad =\sum_{k=1}^3 R_{jk}(g)\bigl[iA_{nn',k}(g^{-1}\bk)\hat\tau_{nn',0}(g) \nonumber\\
  && \quad +\bB_{nn',k}(g^{-1}\bk)\hat{\bm{\tau}}_{nn'}(g)\bigr],
\end{eqnarray}
where $\hat R$ is the $3\times 3$ rotation matrix and
\begin{eqnarray*}
  & \hat\tau_{nn',0}(g)=\hat{\cal D}_n(g)\hat{\cal D}^\dagger_{n'}(g), \\
  & \hat\tau_{nn',\nu}(g)=\hat{\cal D}_n(g)\hat{\sigma}_\nu\hat{\cal D}^\dagger_{n'}(g),\quad \nu=1,2,3.
\end{eqnarray*}
We used the fact that any element of the point group is either a proper rotation $g=R$ or an improper rotation $g=IR$, and that the magnetic field is a pseudovector and not affected by inversion, therefore $g\bH=R(g)\bH$.
The most general phenomenological expressions for the functions $A$ and $\bB$ in a given pair of bands $n$ and $n'$ can be found by solving Eq. (\ref{AB-invariance-g}) for all generators of $\mathbb{G}$, taking into account
the constraints (\ref{A-constraints}) and (\ref{B-constraints}).

\section{Intraband Zeeman coupling}
\label{sec: intraband Zeeman}

We shall see below, in Sec. \ref{sec: SC state}, that the intraband and interband components of the Zeeman coupling produce qualitatively different contributions to the spin magnetic response. Namely, the 
temperature dependence of the spin susceptibility in the superconducting state is almost entirely determined by the intraband coupling. Putting $n=n'$ in Eq. (\ref{A-constraints}), we find $A_{nn,i}=0$. 
Introducing the notation $\bB_{nn,i}=-\bmu_{n,i}$, the spin magnetic moment in the $n$th band can be written as 
\begin{equation}
\label{m_nn}
  \hat m_{nn,i}(\bk)=-\bmu_{n,i}(\bk)\hat{\bm{\sigma}},
\end{equation}
so that the intraband Zeeman coupling takes the following form:
\begin{equation}
\label{Zeeman-intraband}
  \langle\bk,n,s|\hat H_Z|\bk,n,s'\rangle=\sum_{i,\nu}\mu_{n,i\nu}(\bk)H_i\sigma_{\nu,ss'},
\end{equation}
where $i=x,y,z$ and $\nu=1,2,3$.
The coefficients $\mu_{n,i\nu}(\bk)$ form a $3\times 3$ matrix, denoted as $\hat\mu_n(\bk)$, which generalizes the Bohr magneton $\mu_B$ in the $n$th band. 
This matrix is real and even in $\bk$, according to Eq. (\ref{B-constraints}), but not necessarily symmetric, as we shall see below.

From the requirement of invariance under the point group operations, see Eq. (\ref{AB-invariance-g}), we obtain:
\begin{equation}
\label{mu-invariance}
  \hat\mu_n(\bk)=\hat R(g)\hat\mu_n(g^{-1}\bk)\hat{\cal R}^{-1}_n(g),\quad g\in\mathbb{G},
\end{equation}
where the $3\times 3$ orthogonal matrix $\hat{\cal R}_n$ is defined by the following expression:
\begin{equation}
\label{cal R-def}
  \hat{\cal D}_n^\dagger(g)\hat{\sigma}_\mu\hat{\cal D}_n(g)=\sum_{\nu=1}^3{\cal R}_{n,\mu\nu}(g)\hat{\sigma}_\nu.
\end{equation}
This matrix depends on the symmetry of the $n$th band, namely the $\Gamma$-point corep, but not on the band parity. We see that, while the Zeeman coupling matrix $\hat\mu_n(\bk)$ is
invariant under the point group operations, in the sense given by Eq. (\ref{mu-invariance}), it does not transform like a tensor, in general. 

In all pseudospin bands, we have $\hat{\cal D}_n(g)=\hat D^{(1/2)}(R)$ and, using the well-known identity
\begin{equation}
\label{D-to-R}
  \hat{D}^{(1/2),\dagger}(R)\hat{\sigma}_\mu\hat{D}^{(1/2)}(R)=\sum_{\nu}R_{\mu\nu}\hat{\sigma}_\nu,
\end{equation}
we find that $\hat{\cal R}_n(g)=\hat R(g)$ for all elements of the point group. Therefore, Eq. (\ref{mu-invariance}) becomes
\begin{equation}
\label{mu-invariance-pseudospin}
  \hat\mu_n(\bk)=\hat R(g)\hat\mu_n(g^{-1}\bk)\hat R^{-1}(g).
\end{equation}
It is easy to see that the simplest solution of this last equation is given by 
\begin{equation}
\label{usual-Zeeman}
  \mu_{n,i\nu}(\bk)=\mu_n\delta_{i\nu}.
\end{equation}
Thus, in pseudospin bands one can use the ``usual'', i.e., isotropic and momentum-independent, expression for the intraband Zeeman coupling. Note that $\mu_n$ includes the effects of the crystal field and the SO coupling
and therefore does not have to be equal to the Bohr magneton $\mu_B$. In uniaxial crystals, one can use, if needed, a more general solution of Eq. (\ref{mu-invariance-pseudospin}): $\mu_{n,x1}=\mu_{n,y2}=\mu_{n,\perp}$ and 
$\mu_{n,z3}=\mu_{n,z}$, which distinguishes between the basal-plane and $z$-axis Zeeman couplings, but is still $\bk$-independent.

In non-pseudospin bands, using the $\Gamma$-point corep matrices from Table \ref{table: non-pseudospin-corep-matrices}, we obtain that in most cases $\hat{\cal R}_n(g)=\hat R(g)$ and the expression (\ref{usual-Zeeman})
still works. However, in certain bands in trigonal and hexagonal crystals we have $\hat{\cal R}_n(g)\neq\hat R(g)$ for some elements of the point group, so that the invariance equation for $\hat\mu_n(\bk)$ does not have 
the form (\ref{mu-invariance-pseudospin}). Namely, this happens in the following four exceptional cases: 
\begin{equation}
\label{exceptional-bands}
  \begin{array}{l}
  \Gamma_6\ \mathrm{of}\ \mathbf{C}_{3i},\quad (\Gamma_5,\Gamma_6)\ \mathrm{of}\ \mathbf{D}_{3d}, \medskip\\
  (\Gamma_{11},\Gamma_{12})\ \mathrm{of}\ \mathbf{C}_{6h},\quad \Gamma_9\ \mathrm{of}\ \mathbf{D}_{6h}.
  \end{array}
\end{equation}
Note that these are the same bands in which the usual classification of triplet superconducting states fails, see Ref. \onlinecite{Sam19-PRB}.
Below we show that the intraband Zeeman coupling in the exceptional cases is very anisotropic and that some components of $\hat\mu_n(\bk)$ 
necessarily vanish, for symmetry reasons, along the main symmetry axis. The band index $n$ will be dropped, for brevity.

\subsection{$\Gamma_6$ of $\mathbf{C}_{3i}$}
\label{sec: C_3i}

The group $\mathbf{C}_{3i}$ is generated by the rotations $C_{3z}$ about the vertical ($z$) axis and by the inversion $I$. According to Table \ref{table: non-pseudospin-corep-matrices}, 
the $\Gamma$-point corep matrix is given by $\hat{\cal D}_\Gamma(C_{3z})=-\hat\sigma_0$, therefore $\hat{\cal R}(C_{3z})=\hat{\mathbb 1}$ (the $3\times 3$ unit matrix) and the invariance condition (\ref{mu-invariance}) 
takes the following form:
\begin{equation}
\label{mu-equation-C_3i}
  \hat\mu(\bk)=\hat R(C_{3z})\hat\mu(C_{3z}^{-1}\bk).
\end{equation}
Here the rotation matrix is given by 
$$
  \hat R(C_{3z})=\left(\begin{array}{ccc}
                       -1/2 & -\sqrt{3}/2 & 0 \\
                       \sqrt{3}/2 & -1/2 & 0 \\
                       0 & 0 & 1  
                       \end{array}\right)
$$
and the rotations act on the wave vector as $C_{3z}^{-1}k_\pm=e^{\mp 2\pi i/3}k_\pm$ and $C_{3z}^{-1}k_z=k_z$, where $k_\pm=k_x\pm ik_y$. 

The solution of Eq. (\ref{mu-equation-C_3i}) has a block-diagonal form:
\begin{equation}
\label{mu-block-diagonal}
    \hat\mu=\left(\begin{array}{ccc}
                       \mu_{x1} & \mu_{x2} & 0 \\
                       \mu_{y1} & \mu_{y2} & 0 \\
                       0 & 0 & \mu_{z3}
                       \end{array}\right).
\end{equation}
While the equation $\mu_{z3}(\bk)=\mu_{z3}(C_{3z}^{-1}\bk)$ is trivially satisfied by a constant $\mu_{z3}(\bk)=\mu_z$, the solutions for the other components are more involved. Introducing 
$q^\pm_1=\mu_{x1}\pm i\mu_{y1}$ and $q^\pm_2=\mu_{x2}\pm i\mu_{y2}$, we obtain from Eq. (\ref{mu-equation-C_3i}):
\begin{equation}
\label{q1q2-eqs}
  \begin{array}{l}
  q^\pm_1(\bk)=e^{\pm 2\pi i/3}q^\pm_1(C_{3z}^{-1}\bk), \medskip\\ 
  q^\pm_2(\bk)=e^{\pm 2\pi i/3}q^\pm_2(C_{3z}^{-1}\bk),
  \end{array}
\end{equation}
One can easily see that $q^\pm_{1,2}(0,0,k_z)=0$, therefore
\begin{equation}
\label{mu-xy-vanish}
  \mu_{x1}=\mu_{x2}=\mu_{y1}=\mu_{y2}=0\qquad \mathrm{at}\ k_x=k_y=0.
\end{equation}
Thus, the basal-plane components of the Zeeman coupling matrix necessarily depend on $\bk$ and vanish along the threefold symmetry axis.
Using the lowest-order polynomial solutions of Eq. (\ref{q1q2-eqs}), we have
\begin{widetext}
\begin{equation}
\label{mu-C_3i-final}
    \hat\mu(\bk)=\left(\begin{array}{ccc}
                       (\alpha_1k_++\alpha_1^*k_-)k_z & (\alpha_2k_++\alpha_2^*k_-)k_z & 0 \\
                       (-i\alpha_1k_++i\alpha_1^*k_-)k_z & (-i\alpha_2k_++i\alpha_2^*k_-)k_z & 0 \\
                       0 & 0 & \mu_z
                       \end{array}\right),
\end{equation}
\end{widetext}
where $\alpha_{1,2}$ are complex constants and $\mu_z$ is a real constant. Note that the vanishing of the basal-plane components of $\hat\mu$ at $k_z=0$ is accidental, since one can consider a more general solution, 
e.g., $\mu_{x1}=(\alpha_1k_++\alpha_1^*k_-)k_z+(\alpha_3k_+^2+\alpha_3^*k_-^2)$, which has zeros only along the main axis, as dictated by symmetry.

\subsection{$(\Gamma_5,\Gamma_6)$ of $\mathbf{D}_{3d}$}
\label{sec: D_3d}

The group $\mathbf{D}_{3d}$ is generated by the rotations $C_{3z}$ and $C_{2y}$, and also by $I$. From Table \ref{table: non-pseudospin-corep-matrices} we obtain: 
$\hat{\cal R}(C_{3z})=\hat{\mathbb 1}$ and $\hat{\cal R}(C_{2y})=\hat R(C_{2y})$. The first of the two invariance conditions is the same as Eq. (\ref{mu-equation-C_3i}), while the second one has the form 
\begin{equation}
\label{mu-equation-D_3d-C_2y}
    \hat\mu(\bk)=\hat R(C_{2y})\hat\mu(C_{2y}^{-1}\bk)\hat R^{-1}(C_{2y}), 
\end{equation}
where 
$$
  \hat R(C_{2y})=\left(\begin{array}{ccc}
                       -1 & 0 & 0 \\
                       0 & 1 & 0 \\
                       0 & 0 & -1  
                       \end{array}\right)
$$
and $C_{2y}^{-1}(k_+,k_-,k_z)=(-k_-,-k_+,-k_z)$. 

The solution for the Zeeman coupling matrix has a block-diagonal form (\ref{mu-block-diagonal}), with $\mu_{z3}(\bk)=\mu_z$. The condition (\ref{mu-equation-D_3d-C_2y}) produces two additional constraints:
$q^\pm_1(\bk)=q^\mp_1(C_{2y}^{-1}\bk)$ and $q^\pm_2(\bk)=-q^\mp_2(C_{2y}^{-1}\bk)$. Imposing them on Eq. (\ref{mu-C_3i-final}), we obtain:
\begin{equation}
\label{mu-D_3d-final}
    \hat\mu(\bk)=\left(\begin{array}{ccc}
                       \beta_1k_xk_z & -\beta_2k_yk_z & 0 \\
                       \beta_1k_yk_z & \beta_2k_xk_z & 0 \\
                       0 & 0 & \mu_z
                       \end{array}\right),
\end{equation}
where $\beta_{1,2}$ and $\mu_z$ are real constants. The zeros of the basal-plane components of $\hat\mu$ at $k_x=k_y=0$ are required by symmetry, whereas the zero at $k_z=0$ is accidental and can be removed by 
taking into account additional terms in the solution, e.g., $\mu_{x1}=\beta_1k_xk_z+\beta_3(k_x^2-k_y^2)$, with a real $\beta_3$.

\subsection{$(\Gamma_{11},\Gamma_{12})$ of $\mathbf{C}_{6h}$}
\label{sec: C_6h}

The group $\mathbf{C}_{6h}$ is generated by $C_{6z}$ and $I$. From Table \ref{table: non-pseudospin-corep-matrices} we obtain: $\hat{\cal R}(C_{6z})=\hat R(C_{2z})$, therefore the 
invariance condition (\ref{mu-invariance}) takes the following form:
\begin{equation}
\label{mu-equation-C_6h}
  \hat\mu(\bk)=\hat R(C_{6z})\hat\mu(C_{6z}^{-1}\bk)\hat R^{-1}(C_{2z}).
\end{equation}
Here 
\begin{eqnarray*}
  & \hat R(C_{6z})=\left(\begin{array}{ccc}
                       1/2 & -\sqrt{3}/2 & 0 \\
                       \sqrt{3}/2 & 1/2 & 0 \\
                       0 & 0 & 1  
                       \end{array}\right),\\
  & \hat R(C_{2z})=\left(\begin{array}{ccc}
                       -1 & 0 & 0 \\
                       0 & -1 & 0 \\
                       0 & 0 & 1  
                       \end{array}\right),
\end{eqnarray*}
while the action of rotations on the wave vector is given by $C_{6z}^{-1}k_\pm=e^{\mp \pi i/3}k_\pm$ and $C_{6z}^{-1}k_z=k_z$.

The Zeeman coupling matrix has a block-diagonal form (\ref{mu-block-diagonal}), with $\mu_{z3}(\bk)=\mu_z$. The basal-plane components satisfy the equations
$q^\pm_1(\bk)=e^{\mp 2\pi i/3}q^\pm_1(C_{6z}^{-1}\bk)$, $q^\pm_2(\bk)=e^{\mp 2\pi i/3}q^\pm_2(C_{6z}^{-1}\bk)$, and
necessarily vanish along the sixfold symmetry axis. For the lowest-order polynomial solution we obtain:
\begin{equation}
\label{mu-C_6h-final}
    \hat\mu(\bk)=\left(\begin{array}{ccc}
                       \alpha_1k_+^2+\alpha_1^*k_-^2 & \alpha_2k_+^2+\alpha_2^*k_-^2 & 0 \\
                       i\alpha_1k_+^2-i\alpha_1^*k_-^2 & i\alpha_2k_+^2-i\alpha_2^*k_-^2 & 0 \\
                       0 & 0 & \mu_z
                       \end{array}\right),
\end{equation}
where $\alpha_{1,2}$ are complex constants and $\mu_z$ is a real constant.

\subsection{$\Gamma_9$ of $\mathbf{D}_{6h}$}
\label{sec: D_6h}

The group $\mathbf{D}_{6h}$ is generated by the rotations $C_{6z}$ and $C_{2y}$, and by $I$. It follows from Table \ref{table: non-pseudospin-corep-matrices} that $\hat{\cal R}(C_{6z})=\hat R(C_{2z})$ 
and $\hat{\cal R}(C_{2y})=\hat R(C_{2y})$, therefore the invariance conditions are given by Eqs. (\ref{mu-equation-C_6h}) and (\ref{mu-equation-D_3d-C_2y}).
As before, the solution for the Zeeman coupling matrix has a block-diagonal form (\ref{mu-block-diagonal}), where $\mu_{z3}$ is $\bk$-independent, but the basal-plane components necessarily depend on $\bk$.

Imposing the constraint (\ref{mu-equation-D_3d-C_2y}) on Eq. (\ref{mu-C_6h-final}), we obtain:
\begin{equation}
\label{mu-D_6h-final}
    \hat\mu(\bk)=\left(\begin{array}{ccc}
                       \beta_1(k_x^2-k_y^2) & 2\beta_2k_xk_y & 0 \\
                       -2\beta_1k_xk_y & \beta_2(k_x^2-k_y^2) & 0 \\
                       0 & 0 & \mu_z
                       \end{array}\right),
\end{equation}
where $\beta_{1,2}$ and $\mu_z$ are real constants. The vanishing of the basal-plane components along the sixfold axis is required by symmetry.

\subsection{Summary}
\label{sec: intraband-mu-summary}

We have shown that the intraband Zeeman coupling in the four exceptional non-pseudospin bands (\ref{exceptional-bands}) has a nontrivial matrix structure and momentum dependence, see Eqs. (\ref{mu-C_3i-final}), 
(\ref{mu-D_3d-final}), (\ref{mu-C_6h-final}), and (\ref{mu-D_6h-final}). The components of the spin magnetic moment operator in the band representation, Eq. (\ref{m_nn}), have the following form:
\begin{eqnarray}
\label{m_xy-intraband}
  & \hat m_{x,y}(\bk)=\left(\begin{array}{cc}
                          0 & \Phi_{x,y}(\bk)\\
                          \Phi_{x,y}^*(\bk) & 0
                          \end{array}\right),\\
\label{m_z-intraband}                          
  & \hat m_z(\bk)=\left(\begin{array}{cc}
                          -\mu_z & 0\\
                          0 & \mu_z
                          \end{array}\right).
\end{eqnarray}
The basal-plane components $\hat m_{x,y}$ vanish along the main symmetry axis, with the functions $\Phi_x$ and $\Phi_y$ being either linear in $k_x,k_y$ (in trigonal crystals), or quadratic in $k_x,k_y$ (in hexagonal crystals).

\section{Spin susceptibility in the superconducting state}
\label{sec: SC state}

Let us now introduce the superconducting pairing of quasiparticles in the Bloch bands constructed using Eq. (\ref{general-prescription}). Suppose there are $N$ bands crossing the Fermi level and 
participating in superconductivity, then the intraband pairing affected by the Zeeman interaction with an external magnetic field can be described by the following Bardeen-Cooper-Schrieffer (BCS) mean field Hamiltonian:
\begin{equation}
\label{H-general}
  \hat H=\hat H_0+\hat H_Z+\hat H_{sc}.
\end{equation}
The three terms here are as follows: 
\begin{equation}
\label{H_0}
  \hat H_0=\sum_{\bk,n,s}\xi_n(\bk)c^\dagger_{\bk,n,s}c_{\bk,n,s},
\end{equation}
where $\xi_n(\bk)=\xi_n(-\bk)$ are the band dispersions counted from the chemical potential,
\begin{equation}
\label{H-Zeeman-general}
  \hat H_Z=-\sum_{\bk,nn',ss'}\bH\bm{m}_{nn',ss'}(\bk)c^\dagger_{\bk,n,s}c_{\bk,n',s'},
\end{equation}
where the matrix elements of the spin magnetic moment are given by Eq. (\ref{m-matrix elements}), and
\begin{equation}
\label{H-mean-field-general}
  \hat H_{sc}=\frac{1}{2}\sum_{\bk,n,ss'}\left[\Delta_{n,ss'}(\bk)c^\dagger_{\bk,n,s}\tilde c^\dagger_{\bk,n,s'}+\mathrm{H.c.}\right]
\end{equation}
describes the pairing between time-reversed states\cite{And59,Blount85} in the same band, with
\begin{equation}
\label{tilde-c}
  \tilde c^\dagger_{\bk,n,s}=Kc^\dagger_{\bk,n,s}K^{-1}=p_n\sum_{s'}c^\dagger_{-\bk,n,s'}(-i\hat\sigma_2)_{s's},
\end{equation}
according to Eq. (\ref{general-prescription-K}). In the weak-coupling picture, the gap functions are nonzero only inside the BCS shells near the Fermi surfaces: 
$\hat\Delta_n(\bk)\propto\theta(\epsilon_c-|\xi_n(\bk)|)$, where $\epsilon_c$ is the energy cutoff.

By analogy with the standard theory of unconventional superconductivity,\cite{SU-review,TheBook} one can consider separately the pairing in the singlet and triplet channels, keeping in mind that the Kramers index $s$ 
is neither spin nor pseudospin, in general. The matrix structure of the gap functions in the Kramers space can be represented in the following form:
\begin{equation}
\label{Delta-singlet}
  \hat\Delta_n(\bk)=\psi_n(\bk)\hat\sigma_0,\quad \psi_n(-\bk)=\psi_n(\bk),
\end{equation}
in the singlet channel and
\begin{equation}
\label{Delta-triplet}
  \hat\Delta_n(\bk)=\bmd_n(\bk)\hat{\bm{\sigma}},\quad \bmd_n(-\bk)=-\bmd_n(\bk),
\end{equation}
in the triplet channel. Note that, since we defined the gap function in Eq. (\ref{H-mean-field-general}) as a measure of the pairing between the time-reversed states $|\bk,n,s\rangle$ and
$K|\bk,n,s'\rangle$, instead of between $|\bk,n,s\rangle$ and $|-\bk,n,s'\rangle$, the expressions (\ref{Delta-singlet}) and (\ref{Delta-triplet}) do not contain the factors $i\hat\sigma_2$, 
in contrast to the convention used in Refs. \onlinecite{SU-review} and \onlinecite{TheBook}. 

According to Eq. (\ref{general-prescription}), the transformation rule for the electron creation operators in the Bloch states is given by $gc^\dagger_{\bk,n,s}g^{-1}=\sum_{s'}c^\dagger_{g\bk,n,s'}{\cal D}_{n,s's}(g)$. 
Using the fact that the TR operation $K$ commutes with all elements of the point group and is antilinear, we obtain that the point-group rotations and reflections in the physical space induce the following transformation of 
the gap function matrix in the momentum space:\cite{Sam19-PRB} 
$$
  g: \hat\Delta_n(\bk)\to\hat{\cal D}_n(g)\hat\Delta_n(g^{-1}\bk)\hat{\cal D}^\dagger_n(g). 
$$  
In terms of the singlet and triplet components, this becomes
\begin{equation}
\label{singlet-triplet-g}
  g: \psi_n(\bk)\to\psi_n(g^{-1}\bk),\ \bmd_n(\bk)\to{\cal R}_n(g)\bmd_n(g^{-1}\bk),
\end{equation}
where the $3\times 3$ orthogonal matrix $\hat{\cal R}_n$ is defined by Eq. (\ref{cal R-def}). 

It follows from Eq. (\ref{singlet-triplet-g}) that the singlet gap function always transforms under the point group as a complex scalar, regardless of the band symmetry, therefore the usual 
classification of the singlet superconducting states\cite{SU-review,TheBook} is applicable. In contrast, the transformation properties of the triplet gap function essentially depend 
on the band symmetry at the $\Gamma$ point. In a pseudospin band, we have $\hat{\cal R}_n(g)=\hat R$ and $\bmd_n(\bk)$ transforms under $g$ into $R\bmd_n(g^{-1}\bk)$, i.e., like a pseudovector. 
However, this is not the case in general: if the $\Gamma$-point corep is such that $\hat{\cal R}_n(g)\neq\hat R$, which happens in the exceptional bands in trigonal and hexagonal crystals, 
see Eq. (\ref{exceptional-bands}), then $\bmd_n(\bk)$ does not transform under $g$ like a pseudovector, with profound consequences for its momentum dependence and the nodal structure.\cite{Sam19-PRB}

\subsection{Magnetic response}
\label{sec: spin magnetization}

Writing the Zeeman Hamiltonian (\ref{H-Zeeman-general}) in the form $\hat H_Z=-\bH\hat{\cal{\bm M}}$, where $\hat{\cal{\bm M}}$ is the operator of the total spin magnetic moment of electrons, 
we define the magnetization as $\bm{M}={\cal V}^{-1}\langle\hat{\cal{\bm M}}\rangle$ (the angular brackets denote the thermodynamic average and ${\cal V}$ is the volume of the system). 
Introducing the normal and anomalous Green's functions, see Appendix \ref{app: GFs}, and taking the thermodynamic limit ${\cal V}\to\infty$, we obtain:
\begin{equation}
\label{M-from-G}
  \bm{M}=T\sum_m\int\frac{d^3\bk}{(2\pi)^3}\sum_{nn'}\tr[\hat{\bm{m}}_{nn'}(\bk)\hat G_{n'n}(\bk,\omega_m)],
\end{equation}
where $\omega_m=(2m+1)\pi T$ is the fermionic Matsubara frequency and ``$\tr$'' denotes the $2\times 2$ matrix trace with respect to the conjugation indices. We would like to note that the non-pseudospin 
character of the electron bands does not affect the way $\bm{M}$ transforms under the symmetry operations. If the physical state of the system is acted upon by an operation $g\in\mathbb{G}$, then 
$\langle\hat{\cal{\bm M}}\rangle$ is transformed into $\langle g^{-1}\hat{\cal{\bm M}}g\rangle$. Using Eq. (\ref{AB-invariance-g}), it is easy to show that $g^{-1}\hat{\cal M}_ig=\sum_{j}R_{ij}(g)\hat{\cal M}_j$, 
where $R(g)$ is the rotational part of $g$, and therefore $\bm{M}\to R(g)\bm{M}$, as expected.

In a weak field, expanding the Green's function in Eq. (\ref{M-from-G}) in powers of $\bH$, we find $\bm{M}=\bm{M}_0+\bm{M}_1+{\cal O}(H^2)$. The first term here is the spontaneous magnetization given by
\begin{equation}
\label{M_0}
  M_{0,i}=-T\sum_m\int\frac{d^3\bk}{(2\pi)^3}\sum_n\bmu_{n,i}(\bk)\tr[\hat{\bm{\sigma}}\hat G_n(\bk,\omega_m)],
\end{equation}
according to Eq. (\ref{m_nn}), whereas the second term describes the linear magnetic response and can be written as $M_{1,i}=\sum_j\chi^{(spin)}_{ij}H_j$, where
\begin{widetext}
\begin{eqnarray}
\label{chi-general}
  \chi^{(spin)}_{ij} = -T\sum_m\int\frac{d^3\bk}{(2\pi)^3}\sum_{nn'}\tr\bigl[\hat m_{nn',i}(\bk)\hat G_{n'}(\bk,\omega_m)\hat m_{n'n,j}(\bk)\hat G_n(\bk,\omega_m) \nonumber\\
            +\hat m_{nn',i}(\bk)\hat F_{n'}(\bk,\omega_m)\hat m_{n'n,j}(\bk)\hat F^\dagger_n(\bk,\omega_m)\bigr]
\end{eqnarray}
\end{widetext}
is the spin susceptibility tensor. The Green's functions in Eqs. (\ref{M_0}) and (\ref{chi-general}) correspond to zero field and are band-diagonal, see Appendix \ref{app: GFs}.
The expressions (\ref{M_0}) and (\ref{chi-general}) are the multiband generalizations of the standard formulas for the spontaneous magnetization and the spin susceptibility in unconventional superconductors.

In this paper, we consider only unitary superconducting states, which satisfy $\hat\Delta_n\hat\Delta_n^\dagger\propto\hat\sigma_0$. According to Eqs. (\ref{Delta-singlet}) and (\ref{Delta-triplet}), any singlet state is unitary,
whereas a triplet state is unitary only if $\bmd_n\times\bmd_n^*=\bm{0}$. The Green's functions have the following form:
\begin{equation}
\label{GFs-zero-field}
  \begin{array}{l}
    \hat G_n(\bk,\omega_m)=-\dfrac{i\omega_m+\xi_n(\bk)}{\omega_m^2+E_n^2(\bk)}\hat\sigma_0,\medskip \\
    \hat F_n(\bk,\omega_m)=\dfrac{\hat\Delta_n(\bk)}{\omega_m^2+E_n^2(\bk)},
  \end{array}
\end{equation}
where
\begin{equation}
\label{excitation-energy}
  E_n(\bk)=\sqrt{\xi_n^2(\bk)+|\Delta_n(\bk)|^2}
\end{equation}
is the energy of quasiparticle excitations. The energy gap is given by $|\Delta_n|=|\psi_n|$ in the singlet case and $|\Delta_n|=|\bmd_n|$ in the unitary triplet case. The zeros of the energy (the ``gap nodes'') in the $n$th
band correspond to the intersections of the Fermi surface, defined by $\xi_n(\bk)=0$, with the manifold of zeros of $\psi_n(\bk)$ or $\bmd_n(\bk)$. 

We obtain from Eq. (\ref{M_0}) that $\bm{M}_0=\bm{0}$, i.e., there is no spontaneous spin magnetization in a unitary state, regardless 
of the symmetry of the electron bands involved in the pairing. The spin susceptibility tensor (\ref{chi-general}) can be represented in the following form:
\begin{equation}
\label{chi-intra-inter}
    \chi^{(spin)}_{ij}=\tilde\chi_{ij}+\sum_{n}\chi_{n,ij},
\end{equation}
where the first term contains the interband contributions (corresponding to $n\neq n'$) and the second term is the sum of the intraband susceptibilities: 
\begin{eqnarray}
\label{chi-intra}
  \chi_{n,ij}=-T\sum_m\int\frac{d^3\bk}{(2\pi)^3}\tr\bigl[(\bmu_{n,i}\hat{\bm{\sigma}})\hat G_{n}(\bmu_{n,j}\hat{\bm{\sigma}})\hat G_n \nonumber\\
    +(\bmu_{n,i}\hat{\bm{\sigma}})\hat F_{n}(\bmu_{n,j}\hat{\bm{\sigma}})\hat F^\dagger_n\bigr].
\end{eqnarray}
Below we show that the temperature dependence of the susceptibility in the superconducting state is determined by the intraband terms, while the interband contribution is essentially
unaffected by the superconducting transition. 

Let us start with the normal (``$N$'') state. Using the fact that the intraband Zeeman coupling $\hat\mu_n$ is real, we obtain from Eq. (\ref{chi-intra}) the following expression:
\begin{eqnarray}
\label{chi-intra-normal}
  \chi^N_{n,ij}=2\int\frac{d^3\bk}{(2\pi)^3}\left(-\frac{\partial f}{\partial\xi_n}\right)\bigl[\hat\mu_n(\bk)\hat\mu_n^{\dagger}(\bk)\bigr]_{ij} \nonumber\\ 
	      =2N_{F,n}\left\langle\bigl[\hat\mu_n(\bk)\hat\mu_n^{\dagger}(\bk)\bigr]_{ij}\right\rangle_{FS,n},
\end{eqnarray}
where $f(\epsilon)=1/(e^{\epsilon/T}+1)$ is the Fermi function, $N_{F,n}$ is the Fermi-level density of states in the $n$th band, and the angular brackets denote the Fermi-surface average. 
Thus, the intraband susceptibility is determined by the quasiparticles near the Fermi surface and is essentially temperature independent. The ``usual'' expression is recovered 
if one neglects the anisotropy of the Zeeman coupling: putting $\mu_{n,i\nu}(\bk)=\mu_B\delta_{i\nu}$, we obtain: $\chi^N_{n,ij}=\chi_P\delta_{ij}$, where $\chi_P=2N_{F,n}\mu_B^2$ is the Pauli susceptibility. 

Regarding the interband susceptibility, we focus on just one pair of bands ($n,n'=1,2$) with $\xi_2(\bk)>\xi_1(\bk)$, so that the band splitting is given by ${\cal E}(\bk)=\xi_2(\bk)-\xi_1(\bk)$.
Introducing the notations $A_{12,i}=-A_{21,i}=a_i$ and $\bB_{12,i}=\bB_{21,i}=\bmb_i$, see Eqs. (\ref{A-constraints}) and (\ref{B-constraints}), the interband components of the magnetic moment take the following form:
\begin{equation}
\label{m_12}
  \begin{array}{l}
  \hat m_{12,i}(\bk)=ia_i(\bk)+\bmb_i(\bk)\hat{\bm{\sigma}},\medskip\\ 
  \hat m_{21,i}(\bk)=-ia_i(\bk)+\bmb_i(\bk)\hat{\bm{\sigma}}.
  \end{array}
\end{equation}
Here $a_i$ and $\bmb_i$ are real functions of $\bk$, whose parity is determined by the relative parity of the bands. From Eq. (\ref{chi-general}), we obtain:
\begin{equation}
\label{chi-inter-normal}
  \tilde\chi^{N}_{ij}=4\int\frac{d^3\bk}{(2\pi)^3}(a_ia_j+\bmb_i\bmb_j)\frac{f(\xi_1)-f(\xi_2)}{\xi_2-\xi_1}.
\end{equation}
We see that the interband susceptibility is determined by all quasiparticles in the momentum-space shell ``sandwiched'' between the two Fermi surfaces and its temperature dependence is negligible at temperatures 
small compared to the band splitting. 

One can also expect that the interband susceptibility is not affected by the superconducting transition. Indeed, the difference between its values in the superconducting 
(``$S$'') and normal states is very small: even at $T=0$, it can be estimated as $\tilde\chi^{S}_{ij}-\tilde\chi^{N}_{ij}\propto(|\Delta|^2/{\cal E}^2)\ln(\epsilon_c/|\Delta|)$, see Appendix \ref{app: interband chi}. 
In weak-coupling superconductors, the band splitting far exceeds the energy scales associated with superconductivity: $|\Delta_{1,2}|\ll\epsilon_c\ll{\cal E}$, therefore
one can put $\tilde\chi^{S}_{ij}=\tilde\chi^{N}_{ij}$.  

Summarizing our findings so far, the total spin susceptibility (\ref{chi-intra-inter}) can be written in the form
\begin{equation}
\label{chi-intra-inter-final}
    \chi^{(spin)}_{ij}(T)=\tilde\chi_{ij}^N+\sum_{n}\chi_{n,ij}(T),
\end{equation}
where the first term is the temperature-independent interband contribution (\ref{chi-inter-normal}). The second term is given by Eq. (\ref{chi-intra}) and can be calculated in each band individually.

\subsection{Intraband susceptibility}
\label{sec: intraband susceptibility}

The normal and anomalous Green's functions are given by Eq. (\ref{GFs-zero-field}). In a singlet state, dropping the band index $n$ and summing over the Matsubara frequencies in Eq. (\ref{chi-intra}), we obtain:
\begin{equation}
\label{chi-intra-singlet}
  \chi_{ij}(T)=2N_F\left\langle\bigl[\hat\mu(\bk)\hat\mu^\dagger(\bk)\bigr]_{ij}{\cal Y}(\bk,T)\right\rangle_{FS},
\end{equation}
where
\begin{equation}
\label{Yosida-function}
    {\cal Y}(\bk,T)=\frac{1}{2T}\int_0^\infty\frac{d\xi}{\cosh^2(\sqrt{\xi^2+|\Delta(\bk)|^2}/2T)}
\end{equation}
is the angle-resolved Yosida function (Ref. \onlinecite{TheBook}) and $|\Delta(\bk)|=|\psi(\bk)|$ is the energy gap. Being a measure of the number of the thermally excited Bogoliubov quasiparticles 
in the superconducting state, the Yosida function vanishes at $T\to 0$. In the normal state, we have ${\cal Y}=1$ and Eq. (\ref{chi-intra-normal}) is recovered.

In the triplet case, the calculation is somewhat more involved. Introducing the parallel and transverse projectors onto the triplet gap function $\bmd(\bk)$,
$$
  \Pi_{\parallel,\mu\nu}=\frac{\re(d_\mu d_\nu^*)}{|\bmd|^2},\quad \Pi_{\perp,\mu\nu}=\delta_{\mu\nu}-\frac{\re(d_\mu d_\nu^*)}{|\bmd|^2},
$$
we obtain:
\begin{eqnarray}
\label{chi-intra-triplet}
  \chi_{ij}(T)=2N_F\left\langle\bigl[\hat\mu(\bk)\hat \Pi_\parallel(\bk)\hat\mu^\dagger(\bk)\bigr]_{ij}{\cal Y}(\bk,T)\right\rangle_{FS} \nonumber\\
    +2N_F\left\langle\bigl[\hat\mu(\bk)\hat \Pi_\perp(\bk)\hat\mu^\dagger(\bk)\bigr]_{ij}\right\rangle_{FS},
\end{eqnarray}
see Appendix \ref{app: triplet chi} for details. Here ${\cal Y}$ is the Yosida function (\ref{Yosida-function}) with $|\Delta(\bk)|=|\bmd(\bk)|$. 

As explained in Sec. \ref{sec: intraband Zeeman}, in most bands one can use the isotropic Zeeman coupling of the form $\mu_{i\nu}(\bk)=\mu_B\delta_{i\nu}$, see Eq. (\ref{usual-Zeeman}) with the Bohr magneton used for 
simplicity. This leads to the standard expressions for the spin susceptibility. Namely, in the singlet case Eq. (\ref{chi-intra-singlet}) yields $\chi_{ij}(T)=\chi(T)\delta_{ij}$, with
\begin{eqnarray}
\label{chi-intra-singlet-usual}
  \chi(T) &=& \chi_P\left\langle{\cal Y}(\bk,T)\right\rangle_{FS} \nonumber\\
      &=& \chi_P\frac{1}{2T}\int_0^\infty\frac{\nu(E)\,dE}{\cosh^2(E/2T)}.
\end{eqnarray}
We introduced the dimensionless density of states (DoS) of the Bogoliubov quasiparticles: 
\begin{equation}
\label{DoS}
  \nu(E)=\left\langle\frac{E}{\sqrt{E^2-|\Delta(\bk)|^2}}\right\rangle_{FS},
\end{equation}
where the angular integration is performed over the directions satisfying $|\Delta(\bk)|\leq E$. 
Thus, at low temperatures the susceptibility is entirely determined by the thermally excited quasiparticles near the gap nodes and has the following asymptotics:\cite{SU-review,TheBook}
\begin{equation}
\label{chi-singlet-usual-asymptotics}
  \chi(T)\propto
  \left\{ \begin{array}{ll}
         e^{-\Delta/T}, & \mathrm{fully\ gapped},\\
         T^{2/k}, & k^{\mathrm{th}}\ \mathrm{order\ point\ nodes},\\
         T, & \mathrm{line\ nodes}.
         \end{array} \right. 
\end{equation}
Here $\Delta$ denotes the value of the gap in an isotropic superconducting state or the minimum gap in a nodeless anisotropic state. 

In the triplet case with the isotropic Zeeman coupling $\mu_{i\nu}(\bk)=\mu_B\delta_{i\nu}$, we obtain from Eq. (\ref{chi-intra-triplet}):
\begin{eqnarray}
\label{chi-intra-triplet-usual}
  \chi_{ij}(T)=\chi_P\left\langle \Pi_{\parallel,ij}(\bk){\cal Y}(\bk,T)\right\rangle_{FS} \nonumber\\
    +\chi_P\left\langle \Pi_{\perp,ij}(\bk)\right\rangle_{FS},
\end{eqnarray}
i.e., the magnetic response depends on the relative orientation of $\bmd$ and $\bH$ (Ref. \onlinecite{Leggett75}). If $\bmd(\bk)\parallel\bH$, 
then the susceptibility is determined only by the excitations, see the first term in Eq. (\ref{chi-intra-triplet-usual}), and vanishes at $T\to 0$. If $\bmd(\bk)\perp\bH$, then 
both the Cooper pairs and the excitations contribute to the susceptibility, see the second term in Eq. (\ref{chi-intra-triplet-usual}), leading to a nonzero residual susceptibility at $T=0$.

In the exceptional bands, the intraband Zeeman coupling has a nontrivial matrix structure and a complicated momentum dependence, see Sec. \ref{sec: intraband Zeeman}. 
Therefore, Eqs. (\ref{chi-intra-singlet-usual}) and (\ref{chi-intra-triplet-usual}) are no longer applicable. In particular, the quasiparticle contribution to the susceptibility is determined not only by the gap nodal structure 
through the quasiparticle DoS (\ref{DoS}), but also by the zeros of the Zeeman coupling. Namely, if the nodal excitations are weakly coupled with the external magnetic field, then they do not contribute to the magnetization, 
which leads to significant changes in the temperature dependence of $\chi_{ij}$.

\section{Application to hexagonal superconductors}
\label{sec: hexagonal}

As an application of the general theory developed above, in this section we calculate the spin susceptibility of a hexagonal superconductor described by the point group $\mathbf{D}_{6h}=\mathbf{D}_6\times\{E,I\}$. 
This group has six double-valued coreps of even or odd parity,\cite{Lax-book} corresponding to the $\Gamma^\pm_7$, $\Gamma^\pm_8$, or $\Gamma^\pm_9$ bands, with only the $\Gamma^+_7$ bands being pseudospin ones.
The parity superscript can be omitted, because neither the intraband pairing nor the intraband Zeeman coupling are affected by the band parity. 
We assume that there is just one isotropic band of $\Gamma_7$, $\Gamma_8$, or $\Gamma_9$ symmetry crossing the Fermi level and participating in superconductivity.

According to the Landau theory of phase transitions, the superconducting gap functions are classified according to single-valued irreducible representations (irreps) $\gamma$ of the point group $\mathbb{G}$, 
called the pairing channels. In a $d$-dimensional pairing channel, the singlet gap function can be represented in the form $\psi(\bk)=\sum_{a=1}^{d}\eta_a\varphi_a(\bk)$,
where $a$ labels the scalar basis functions $\varphi$ of an even irrep $\gamma$. Similarly, the triplet gap function can be represented as $\bmd(\bk)=\sum_{a=1}^{d}\eta_a\bm{\varphi}_a(\bk)$,
where the basis functions $\bm{\varphi}$ of an odd irrep $\gamma$ have different form in the pseudospin and non-pseudospin bands, see Ref. \onlinecite{Sam19-PRB}. 
In both singlet and triplet cases, the components of the superconducting order parameter $\eta_1,..,\eta_d$ are found by minimizing the free energy of the superconductor.\cite{TheBook,SU-review}
The point group $\mathbf{D}_6$ has four one-dimensional (1D) irreps $A_1$, $A_2$, $B_1$, $B_2$, and two 2D irreps $E_1$, $E_2$ (we use the ``chemical'' notation\cite{LL-3} for the pairing channels, 
reserving the $\Gamma$ notation\cite{Lax-book} for the double-valued coreps describing the symmetry of the Bloch bands). Taking into account the parity, there are six singlet and six triplet pairing channels.

\subsection{Singlet pairing}
\label{sec: D_6h-singlet}

Examples of the basis functions for the even irreps of $\mathbf{D}_{6h}$ are shown in Table \ref{table: D_6h-singlet}.
The basis functions in all channels except $A_{1g}$ (which corresponds to the trivial, or ``$s$-wave'', pairing) have zeros imposed by symmetry. The $A_{2g}$ gap has twelve vertical lines of nodes, whereas the $B_{1g}$ and $B_{2g}$
gaps have six vertical lines of nodes as well as a horizontal line of nodes at $k_z=0$.
For the 2D pairing channels, the nodal structure depends on the state, i.e., on the order parameter components $(\eta_1,\eta_2)$. The stable states correspond to $(\eta_1,\eta_2)\propto (1,0)$ or $(1,1)$ 
(Ref. \onlinecite{SU-review}).

\begin{table}
\caption{Examples of the basis functions $\varphi(\bk)$ in the even pairing channels for $\mathbb{G}=\mathbf{D}_{6h}$ ($a$ is a real constant and $k_\pm=k_x\pm ik_y$).}
{\begin{tabular}{ll}
    \toprule
    $\gamma$ &  $\Gamma_7$, $\Gamma_8$, and $\Gamma_9$ bands \\ \colrule
    $A_{1g}$   &  $k_x^2+k_y^2+ak_z^2$  \\ \colrule
    $A_{2g}$   &  $i(k_+^6-k_-^6)$  \\ \colrule
    $B_{1g}$   &  $(k_+^3+k_-^3)k_z$   \\ \colrule
    $B_{2g}$   &  $i(k_+^3-k_-^3)k_z$  \\ \colrule
    $E_{1g}$   &  $k_+k_z$, $k_-k_z$  \\ \colrule
    $E_{2g}$\hspace*{4mm}   &  $k_+^2$, $k_-^2$ \\ \botrule
\end{tabular}}
\label{table: D_6h-singlet}
\end{table}

\subsubsection{$\Gamma_7$ and $\Gamma_8$ bands}
\label{sec: D_6h-singlet-G7G8}

In the non-exceptional $\Gamma_7$ and $\Gamma_8$ bands, one can use the usual expression for the Zeeman coupling, $\mu_{i\nu}(\bk)=\mu_B\delta_{i\nu}$. The susceptibility is determined by the quasiparticle DoS, 
see Eq. (\ref{chi-intra-singlet-usual}), with the standard low-temperature asymptotics (\ref{chi-singlet-usual-asymptotics}).

\subsubsection{$\Gamma_9$ bands}
\label{sec: D_6h-singlet-G9}

Substituting the Zeeman coupling (\ref{mu-D_6h-final}) into Eq. (\ref{chi-intra-singlet}) and using the fact that the Yosida function has the same symmetry as the energy gap, 
one can show that the off-diagonal elements of the susceptibility tensor vanish after the Fermi-surface averaging, whereas the diagonal elements have the following form:
\begin{eqnarray}
\label{chi-hexagonal-singlet-G9}
  && \chi_{xx}=2N_F\bigl\langle \bigl[\beta_1^2(k_x^2-k_y^2)^2+4\beta_2^2k_x^2k_y^2\bigr] \nonumber\\
  &&    \hspace*{4.2cm}\times{\cal Y}(\bk,T) \bigr\rangle_{FS}, \nonumber \\
  && \chi_{yy}=2N_F\bigl\langle \bigl[4\beta_1^2k_x^2k_y^2+\beta_2^2(k_x^2-k_y^2)^2\bigr] \\
  &&    \hspace*{4.2cm}\times{\cal Y}(\bk,T) \bigr\rangle_{FS}, \nonumber\\
  && \chi_{zz}=2N_F\mu_z^2\left\langle {\cal Y}(\bk,T) \right\rangle_{FS}. \nonumber
\end{eqnarray}
In the normal state, we have ${\cal Y}=1$ and the susceptibility components are given by
\begin{equation}
 \label{chi-hexagonal-normal-G9}
 \begin{array}{l}
  \chi_{xx}^N=\chi_{yy}^N=\dfrac{8}{15}N_Fk_F^4(\beta_1^2+\beta_2^2)\equiv\chi_\perp^N,\medskip \\
  \chi_{zz}^N=2N_F\mu_z^2,
 \end{array}
\end{equation}
assuming a spherical Fermi surface of radius $k_F$. In all superconducting states, the temperature dependence of $\chi_{zz}$ is entirely determined by the quasiparticle DoS, 
see Eq. (\ref{chi-intra-singlet-usual}), and has the asymptotics (\ref{chi-singlet-usual-asymptotics}). 

In contrast, the basal-plane components $\chi_{xx}$ and $\chi_{yy}$ contain the Yosida function multiplied by strongly anisotropic factors, which vanish along the sixfold symmetry axis, 
therefore the contribution of the quasiparticles near the main axis to the Fermi-surface averages is suppressed. The effect of this suppression is most pronounced in the superconducting states which 
have only isolated point nodes at $k_x=k_y=0$. Using Table \ref{table: D_6h-singlet}, it is easy to see that there is only one such singlet state, namely, the ``chiral'' $d$-wave state of the form 
$\psi(\bk)\propto k_+^2$ (or $k_-^2$),
which corresponds to the $E_{2g}$ irrep and has isolated second-order point nodes along the main axis. In this state, the expressions (\ref{chi-hexagonal-singlet-G9}) yield
\begin{equation}
\label{chi-E2g-G9}
    \chi_{xx}(T)=\chi_{yy}(T)\propto T^3,\quad \chi_{zz}(T)\propto T, 
\end{equation}
at $T\to 0$. In all other singlet states, $\chi_{xx}$ and $\chi_{yy}$ have the asymptotics (\ref{chi-singlet-usual-asymptotics}).

\subsection{Triplet pairing}
\label{sec: D_6h-triplet}

In the triplet case, the gap function $\bmd(\bk)$ and the basis functions $\bm{\varphi}(\bk)$ can be represented as linear combinations of three orthonormal pseudovectors $\be_1$, $\be_2$, and $\be_3$, whose 
transformations induced by the point group operations follow from Eq. (\ref{singlet-triplet-g}), namely, $\be_\mu\to {\cal R}(g)\be_\mu=\sum_\nu\be_\nu{\cal R}_{\nu\mu}(g)$.
Note that $\be_1$, $\be_2$, $\be_3$ are not the same as the Cartesian basis vectors $\hbx$, $\hby$, $\hbz$ in the physical space, which is easily seen from their different response to inversion.  
However, they are not entirely independent: if the spin quantization axis is chosen along $\hbz$, then all corep matrices for the rotations about $\hbz$, see Table \ref{table: non-pseudospin-corep-matrices}, 
commute with $\hat\sigma_3$ and we obtain from Eq. (\ref{cal R-def}) that $\be_3$ is unchanged by these rotations. It is in this sense that $\be_3$ is ``parallel'' to $\hbz$. 

The triplet basis functions for the odd pairing channels are shown in Table \ref{table: D_6h-triplet}. In the $\Gamma_7$ and $\Gamma_8$ bands we have $\hat{\cal R}(g)=\hat R$ for all $g$, therefore $\bmd$ and $\be_\mu$ 
transform under rotations like vectors and the standard expressions for the basis functions\cite{SU-review,TheBook} are applicable. In contrast, in the $\Gamma_9$ bands we have $\hat{\cal R}(g)\neq\hat R$ for some $g$, 
therefore $\bmd$ and $\be_\mu$ do not transform under all rotations like vectors, which changes the momentum dependence of the basis functions.\cite{Sam19-PRB} Note that all three components of $\bmd$ never
vanish simultaneously in a whole plane in the momentum space, so that Blount's theorem about the absence of line nodes in a generic triplet state\cite{Blount85} holds in the exceptional non-pseudospin bands as well.

The precise form of the basis functions in a given material, in particular, the relative values of the coefficients in Table \ref{table: D_6h-triplet}, is dictated by the microscopic details. 
We will consider two cases: $\bmd\parallel\hbz$, which corresponds to $a_1=a_2=0$ in all basis functions, and $\bmd\perp\hbz$. In both cases, the triplet states have isolated point nodes and/or accidental lines of nodes.

\begin{table*}
\caption{Examples of the basis functions $\bm{\varphi}(\bk)$ in the odd pairing channels for $\mathbb{G}=\mathbf{D}_{6h}$, where $a_{1,2,3}$ are real constants, $k_\pm=k_x\pm ik_y$, and $\be_\pm=(\be_1\pm i\be_2)/\sqrt{2}$. 
Second column: the bands in which $\bmd$ transforms under rotations like a vector. Third column: the bands in which $\bmd$ does not transform like a vector.}
{\begin{tabular}{lll}
    \toprule
    $\gamma$ & $\Gamma_7$ and $\Gamma_8$ bands & $\Gamma_9$ bands  \\ \colrule
    $A_{1u}$   &  $a_1(k_x\be_1+k_y\be_2)+a_3k_z\be_3$  &  $a_1(k_+^3+k_-^3)\be_1+ia_2(k_+^3-k_-^3)\be_2+a_3k_z\be_3$ \\ \colrule
    $A_{2u}$   &  $a_1(k_y\be_1-k_x\be_2)+ia_3(k_+^6-k_-^6)k_z\be_3$  &  $ia_1(k_+^3-k_-^3)\be_1+a_2(k_+^3+k_-^3)\be_2+ia_3(k_+^6-k_-^6)k_z\be_3$  \\ \colrule
    $B_{1u}$   &  $a_1(k_+^2k_z\be_++k_-^2k_z\be_-)+a_3(k_+^3+k_-^3)\be_3$  &  $a_1k_z\be_1+ia_2(k_+^6-k_-^6)k_z\be_2+a_3(k_+^3+k_-^3)\be_3$  \\ \colrule
    $B_{2u}$   &  $ia_1(k_+^2k_z\be_+-k_-^2k_z\be_-)+ia_3(k_+^3-k_-^3)\be_3$\hspace*{4mm}   &  $ia_1(k_+^6-k_-^6)k_z\be_1+a_2k_z\be_2+ia_3(k_+^3-k_-^3)\be_3$ \\ \colrule
    $E_{1u}$   &  $a_1k_z\be_++a_3k_+\be_3$,  &   $a_1k_-^2k_z\be_1+a_2k_-^2k_z\be_2+a_3k_+\be_3$, \\
               &     $a_1k_z\be_-+a_3k_-\be_3$   & $a_1k_+^2k_z\be_1-a_2k_+^2k_z\be_2+a_3k_-\be_3$ \\ \colrule
    $E_{2u}$\hspace*{4mm}   &  $a_1k_+\be_++a_3k_+^2k_z\be_3$, &  $a_1k_+\be_1+a_2k_+\be_2+a_3k_-^2k_z\be_3$,\\
               &  $a_1k_-\be_-+a_3k_-^2k_z\be_3$   & $a_1k_-\be_1-a_2k_-\be_2+a_3k_+^2k_z\be_3$ \\ \botrule
\end{tabular}}
\label{table: D_6h-triplet}
\end{table*}

\subsubsection{$\Gamma_7$ and $\Gamma_8$ bands}
\label{sec: D_6h-triplet-G7G8}

In the non-exceptional $\Gamma_7$ and $\Gamma_8$ bands, one can put $\mu_{i\nu}(\bk)=\mu_B\delta_{i\nu}$. Therefore, the susceptibility is given by Eq. (\ref{chi-intra-triplet-usual}) and 
has the usual temperature dependence. For instance, if $\bmd\parallel\hbz$, then $\chi_{xx}=\chi_{yy}=\chi_P$ and $\chi_{zz}=\chi_P\left\langle{\cal Y}(\bk,T)\right\rangle_{FS}$, i.e., the basal-plane susceptibility 
is the same as in the normal state, whereas the $z$-axis susceptibility is determined by the nodal quasiparticles and has the asymptotics (\ref{chi-singlet-usual-asymptotics}) at $T\to 0$.

\subsubsection{$\Gamma_9$ bands}
\label{sec: D_6h-triplet-G9}

The Zeeman coupling in the $\Gamma_9$ bands has the form (\ref{mu-D_6h-final}). If $\bmd\parallel\hbz$, then we obtain from Eq. (\ref{chi-intra-triplet}) that the off-diagonal components vanish after the Fermi-surface averaging,
the basal-plane susceptibility is not affected by the superconductivity, and the $z$-axis susceptibility is determined by the nodal quasiparticles:
\begin{equation}
\label{chi-hexagonal-triplet}
  \begin{array}{c}
  \chi_{xx}(T)=\chi_{yy}(T)=\chi_\perp^N, \medskip\\ 
  \chi_{zz}(T)=\chi_{zz}^N\left\langle{\cal Y}(\bk,T)\right\rangle_{FS},
  \end{array}
\end{equation}
where $\chi_\perp^N$ and $\chi_{zz}^N$ are given by Eq. (\ref{chi-hexagonal-normal-G9}). The low-temperature asymptotics of $\chi_{zz}$ are the same as in Eq. (\ref{chi-singlet-usual-asymptotics}). 
Therefore, the non-pseudospin character of the $\Gamma_9$ bands does not qualitatively affect the spin response for $\bmd\parallel\hbz$. 

In the case $\bmd\perp\hbz$, let us first put $a_2=a_3=0$ in all basis functions in the third column of Table \ref{table: D_6h-triplet}, which corresponds to $\bmd\parallel\hbx$. In all stable states, the 
off-diagonal elements of the susceptibility tensor vanish and we obtain:
\begin{eqnarray}
\label{chi-hexagonal-triplet-G9}
  && \chi_{xx}=2N_F\bigl[\beta_1^2\left\langle(k_x^2-k_y^2)^2{\cal Y}(\bk,T)\right\rangle_{FS} \nonumber\\
  &&    \hspace*{3.4cm} +4\beta_2^2\left\langle k_x^2k_y^2\right\rangle_{FS}\bigr], \nonumber \\
  && \chi_{yy}=2N_F\bigl[4\beta_1^2\left\langle k_x^2k_y^2{\cal Y}(\bk,T)\right\rangle_{FS} \\ 
  &&    \hspace*{2.7cm} +\beta_2^2\left\langle(k_x^2-k_y^2)^2\right\rangle_{FS}\bigr], \nonumber\\
  && \chi_{zz}=2N_F\mu_z^2. \nonumber
\end{eqnarray}
We see that, while $\chi_{zz}$ is not affected by the superconductivity: $\chi_{zz}(T)=\chi_{zz}^N$, the basal-plane components are suppressed in the superconducting state, but not completely, 
as both $\chi_{xx}$ and $\chi_{yy}$ attain a nonzero value at $T=0$:
\begin{equation}
\label{d-along-x-residual-chi}
  \chi_{xx}(T=0)=\chi_{yy}(T=0)=\frac{\beta_2^2}{\beta_1^2+\beta_2^2}\chi_\perp^N.
\end{equation}
In contrast, in the case of the usual isotropic Zeeman coupling one has $\chi_{xx}=\chi_P\left\langle{\cal Y}(\bk,T)\right\rangle_{FS}$ and $\chi_{yy}=\chi_{zz}=\chi_P$.

The temperature dependence of $\chi_{xx}$ and $\chi_{yy}$ is determined by the Fermi-surface averages in the first terms in 
Eq. (\ref{chi-hexagonal-triplet-G9}). Since the Yosida functions are multiplied by strongly anisotropic factors, which vanish in the vertical planes $|k_x|=|k_y|$ (in $\chi_{xx}$) or $k_xk_y=0$ (in $\chi_{yy}$), 
the quasiparticle contribution to the susceptibility is suppressed if the superconducting gap has zeros only in these planes. Using Table \ref{table: D_6h-triplet}, 
it is easy to see that this happens in the non-chiral ``nematic'' $p$-wave state $\bmd(\bk)\propto k_x\be_1$ (or $k_y\be_1$), which corresponds to the $E_{2u}$ irrep and has vertical line nodes. In this state, we obtain:
\begin{equation}
\label{chi-E2u-nonchiral-G9}
    \chi_{xx}(T)-\chi_{xx}(0)\propto T,\quad \chi_{yy}(T)-\chi_{yy}(0)\propto T^3. 
\end{equation}
In the chiral $p$-wave state $\bmd(\bk)\propto k_+\be_1$ (or $k_-\be_1$), the gap has isolated first-order point nodes along the main axis. The anisotropic Zeeman factors in the first terms in $\chi_{xx}$ and $\chi_{yy}$ vanish 
along this axis and we obtain:
\begin{equation}
\label{chi-E2u-chiral-G9}
    \chi_{xx}(T)-\chi_{xx}(0)=\chi_{yy}(T)-\chi_{yy}(0)\propto T^6. 
\end{equation}
We see that the spin susceptibility in both chiral and nonchiral $E_{2u}$ states in the $\Gamma_9$ bands exhibits the temperature dependence which is never found in the $\Gamma_7$ and $\Gamma_8$ bands.

In a more general state with $\bmd\perp\hbz$, the direction of $\bmd$ may change as a function of momentum and the expressions (\ref{chi-hexagonal-triplet-G9}) are no longer applicable. As an example,
let us consider an $f$-wave $A_{2u}$ state with $a_1=a_2\neq 0$ and $a_3=0$, see Table \ref{table: D_6h-triplet}. The gap function
\begin{equation}
\label{A-2u-G9}
  \bmd(\bk)\propto i(k_+^3-k_-^3)\be_1+(k_+^3+k_-^3)\be_2
\end{equation}
has two isolated third-order point nodes at the main symmetry axis. From Eq. (\ref{chi-intra-triplet}), we obtain the following nonzero components of the susceptibility tensor:
\begin{eqnarray}
\label{chi-hexagonal-A-2u-G9}
  && \chi_{xx}=N_F\bigl\langle \bigl[\beta_1^2(k_x^2-k_y^2)^2+4\beta_2^2k_x^2k_y^2\bigr] \nonumber\\
  &&		\hspace*{3.1cm} \times[1+{\cal Y}(\bk,T)]\bigr\rangle_{FS}, \nonumber \\
  && \chi_{yy}=N_F\bigl\langle \bigl[4\beta_1^2k_x^2k_y^2+\beta_2^2(k_x^2-k_y^2)^2\bigr] \\
  &&		\hspace*{3.1cm} \times[1+{\cal Y}(\bk,T)]\bigr\rangle_{FS}, \nonumber\\
  && \chi_{zz}=2N_F\mu_z^2. \nonumber
\end{eqnarray}
Therefore, $\chi_{zz}$ is the same as in the normal state, whereas the basal-plane components are suppressed in the superconducting state, but have a residual value at $T=0$:
\begin{equation}
\label{A-3u-inplane-residual}
  \chi_{xx}(T=0)=\chi_{yy}(T=0)=\frac{1}{2}\chi_\perp^N.
\end{equation}
At nonzero temperatures, we calculate the Fermi-surface averages in Eq. (\ref{chi-hexagonal-A-2u-G9}) containing the Yosida function and obtain:
\begin{equation}
\label{chi-A2u-G9}
    \chi_{xx}(T)-\chi_{xx}(0)=\chi_{yy}(T)-\chi_{yy}(0)\propto T^2,
\end{equation}
which is very different from the $T^{2/3}$ behaviour expected for isolated third-order nodes in the isotropic Zeeman coupling case, see Eq. (\ref{chi-singlet-usual-asymptotics}).

\subsection{Discussion}
\label{sec: UPt3-discussion}

The point group $\mathbf{D}_{6h}$ describes, for instance, the symmetry of the heavy-fermion material UPt$_3$ (Ref. \onlinecite{UPt3-review}), in which a variety of thermodynamic and transport measurements have revealed an 
unconventional superconducting state. Although there is still no consensus on the pairing symmetry in UPt$_3$, the most promising candidate model is based on the 2D irrep $E_{2u}$ of $\mathbf{D}_{6h}$. 
The corresponding order parameter is real in the high-temperature $A$ phase and complex (TR symmetry breaking) in the low-temperature $B$ phase.\cite{Strand09,Schemm14,Avers20} 
The electron bands in UPt$_3$ originate from
the $j_z=\pm 1/2$, $\pm 3/2$, and $\pm 5/2$ doublets at the $\Gamma$ point, which correspond, respectively, to the $\Gamma_7$, $\Gamma_9$, and $\Gamma_8$ coreps of $\mathbf{D}_{6h}$. 

Unfortunately, the NMR Knight shift in UPt$_3$ changes very little in the superconducting state for all field orientations and the experimental results ``are not conclusive in favor of any theory'' 
(Ref. \onlinecite{UPt3-review}), see also Ref. \onlinecite{Gannon17}. One possible expalanation is that in this material there are five bands crossing the Fermi level, which alone would lead 
to a significant temperature-independent interband contribution, see Eq. (\ref{chi-intra-inter-final}), regardless of the pairing symmetry. Additional contribution to the residual susceptibility can come, 
e.g., from impurity scattering.\cite{AG62}

\section{Conclusions}
\label{sec: Conclusion}

We have derived the general symmetry-constrained expressions for the Zeeman coupling in crystals with the SO coupling. We showed that in some Bloch bands in trigonal and hexagonal crystals the Zeeman interaction of
the band electrons with a magnetic field directed in the basal plane necessarily vanishes along the main symmetry axis, either linearly or quadratically in $k_x$ and $k_y$. This significantly changes the low-temperature 
behaviour of the spin susceptibility in the superconducting states with the gap nodes. For example, in a hexagonal crystal, the basal-plane intraband susceptibility in the $\Gamma_9$ bands has the form 
$\chi_{xx}(T)=\chi_{yy}(T)\propto T^3$ in the singlet chiral $d$-wave state (instead of the ``usual'' linear in $T$ dependence, for the isotropic Zeeman coupling). In the triplet states, one has  
$\chi_{xx}(T)-\chi_{xx}(0)\propto T$, $\chi_{yy}(T)-\chi_{yy}(0)\propto T^3$ in the nematic $p$-wave state 
(instead of the linear in $T$ dependence for $\chi_{xx}$ and a temperature-independent $\chi_{yy}$), and 
$\chi_{xx}(T)-\chi_{xx}(0)=\chi_{yy}(T)-\chi_{yy}(0)\propto T^6$ 
in the chiral $p$-wave state (instead of the $T^2$ dependence for $\chi_{xx}$ and a temperature-independent $\chi_{yy}$).

\acknowledgments

The author is grateful to V. P. Mineev for a stimulating correspondence.
This work was supported by a Discovery Grant 2015-06656 from the Natural Sciences and Engineering Research Council of Canada.

\appendix

\section{Symmetry of the Bloch states}
\label{app: Bloch bases}

The Bloch states $|\bk,n,1\rangle$ and $|\bk,n,2\rangle$ in a crystal with the SO coupling form the basis of an irreducible double-valued corep of the magnetic point group of the wave vector $\bk$. 
The full symmetry group of $\bk$ is ``magnetic'', because it contains the antiunitary conjugation operation ${\cal C}$. A detailed review of magnetic groups and their coreps can be found, for instance, in 
Refs. \onlinecite{BD68} and \onlinecite{BC-book}. If the crystal point group is $\mathbb{G}$, then the magnetic group at the $\Gamma$ point is ${\cal G}=\mathbb{G}+{\cal C}\mathbb{G}$. 
Since we consider only crystals with a center of inversion, the coreps of ${\cal G}$ have a definite parity, i.e., are either inversion-even ($\Gamma^+$) or inversion-odd ($\Gamma^-$). 

The double-valued coreps of ${\cal G}$ can be obtained from the double-valued irreducible representations (irreps) of $\mathbb{G}$, with the results listed in Table \ref{table: Gamma-point-coreps}.
All the coreps are 2D, except $(\Gamma_6^\pm,\Gamma_7^\pm)$ for $\mathbb{G}=\mathbf{T}_{h}$ and $\Gamma_8^\pm$ for $\mathbb{G}=\mathbf{O}_{h}$, 
which are four-dimensional. These latter coreps correspond to the bands which are fourfold degenerate at the $\Gamma$ point, e.g., the $\Gamma_8^\pm$ (``$j=3/2$'') bands 
in cubic crystals (Ref. \onlinecite{Lutt56}), and are not considered here. Pairs of complex conjugate irreps $(\Gamma,\Gamma^*)$ produce coreps of the ``pairing'' type of twice the dimension, 
while the 1D irreps $\Gamma_2^\pm$ of $\mathbf{C}_{i}$ and $\Gamma_6^\pm$ of $\mathbf{C}_{3i}$ produce 2D coreps of the ``doubling'' type.\cite{Lax-book} 

The fourth and fifth columns in Table \ref{table: Gamma-point-coreps} show the spinor basis functions $(\phi_1,\phi_2)$ which reproduce the corep matrices in Table \ref{table: non-pseudospin-corep-matrices}. 
For the pseudospin coreps, the basis functions can be chosen to be the pure spin states $\spup$ and $\spdown$ (the eigenstates of the spin operator $\hat s_z$), whereas the basis functions for the non-pseudospin coreps 
contain the pure spin states multiplied by additional non-removable coordinate-dependent factors. The phases are chosen to satisfy the conjugation condition $\phi_2={\cal C}\phi_1$. 

Transformation of the Bloch states at the $\Gamma$ point in a twofold degenerate band corresponding to a 2D corep ${\cal D}_n$ is given by
\begin{equation}
\label{Bloch-states-transform-Gamma-point}
  g|\bm{0},n,s\rangle=\sum_{s'}|\bm{0},n,s'\rangle {\cal D}_{n,s's}(g).
\end{equation}
The explicit form of the corep matrices depends on the choice of the basis: an arbitrary unitary rotation of the basis, $|\bm{0},n,s\rangle\to|\bm{0},n,s\rangle'=\sum_{s_1}|\bm{0},n,s_1\rangle U_{n,s_1s}$, produces an 
equivalent corep with $\hat{\cal D}'_n(g)=\hat U_n^{-1}\hat{\cal D}_n(g)\hat U_n$ and $\hat{\cal D}'_n({\cal C})=\hat U_n^{-1}\hat{\cal D}_n({\cal C})\hat U_n^*$. The ``orientation'' and the phases 
of the Bloch basis at the $\Gamma$-point can always be chosen to reproduce both the matrices in Table \ref{table: non-pseudospin-corep-matrices} and the matrix representation of the conjugation, 
$\hat{\cal D}_n({\cal C})=-i\hat\sigma_2$.
The prescription (\ref{general-prescription}) is obtained from Eq. (\ref{Bloch-states-transform-Gamma-point}) by using the fact that the Bloch states in the bands that are only twofold degenerate due to the conjugation symmetry
are analytic functions of $\bk$, at least in the vicinity of the $\Gamma$ point, which can be shown, e.g., using the $\bk\cdot\bm{p}$ perturbation theory.

\begin{table}
\caption{The double-valued coreps of the centrosymmetric magnetic point groups at the $\Gamma$ point, with examples of even and odd spinor basis functions for the 2D coreps;  
$f_\Gamma(\br)$ is a real basis function of a single-valued irrep $\Gamma$ of $\mathbb{G}$, and $\rho_\pm=x\pm iy$.}
{\begin{tabular}{lllll}
    \toprule
    $\mathbb{G}$ & corep & $\dim$\hspace*{1mm} & even basis & odd basis \\ \hline
    $\mathbf{C}_{i}$  & $\Gamma_2$ & 2 & $(\spup,\spdown)$  & $if_{\Gamma_1^-}(\br)(\spup,\spdown)$ \\ \hline
    $\mathbf{C}_{2h}$ & $(\Gamma_3,\Gamma_4)$ & 2 & $(\spup,\spdown)$ & $if_{\Gamma_1^-}(\br)(\spup,\spdown)$ \\ \hline
    $\mathbf{D}_{2h}$ & $\Gamma_5$  & 2 & $(\spup,\spdown)$ & $if_{\Gamma_1^-}(\br)(\spup,\spdown)$ \\ \hline
    $\mathbf{C}_{4h}$ & $(\Gamma_5,\Gamma_6)$ & 2 & $(\spup,\spdown)$ & $if_{\Gamma_1^-}(\br)(\spup,\spdown)$ \\ 
                      & $(\Gamma_7,\Gamma_8)$ & 2 & $(\rho_-^2\spup,\rho_+^2\spdown)$ & $(\rho_+\spup,-\rho_-\spdown)$ \\ \hline
    $\mathbf{D}_{4h}$ & $\Gamma_6$  & 2 & $(\spup,\spdown)$ & $if_{\Gamma_1^-}(\br)(\spup,\spdown)$ \\ 
		      & $\Gamma_7$  & 2 &  $(\rho_-^2\spup,\rho_+^2\spdown)$ & $(\rho_+\spup,-\rho_-\spdown)$ \\ \hline
    $\mathbf{C}_{3i}$ & $(\Gamma_4,\Gamma_5)$ & 2 & $(\spup,\spdown)$ & $if_{\Gamma_1^-}(\br)(\spup,\spdown)$ \\ 
                      & $\Gamma_6$ & 2 & $(\rho_-^2\spup,\rho_+^2\spdown)$ &  $(\rho_+\spup,-\rho_-\spdown)$ \\ \hline		       
    $\mathbf{D}_{3d}$ & $\Gamma_4$  & 2 & $(\spup,\spdown)$ & $if_{\Gamma_1^-}(\br)(\spup,\spdown)$ \\ 
		      & $(\Gamma_5,\Gamma_6)$  & 2 & $(\rho_-^2\spup,\rho_+^2\spdown)$ &  $(\rho_+\spup,-\rho_-\spdown)$ \\ \hline
    $\mathbf{C}_{6h}$ & $(\Gamma_7,\Gamma_8)$  & 2 & $(\spup,\spdown)$ & $if_{\Gamma_1^-}(\br)(\spup,\spdown)$ \\ 
		      & $(\Gamma_9,\Gamma_{10})$  & 2 & $(\rho_+^2\spup,\rho_-^2\spdown)$  & $(\rho_-^3\spup,-\rho_+^3\spdown)$ \\
                      & $(\Gamma_{11},\Gamma_{12})$\hspace*{2mm} & 2 &  $(\rho_-^2\spup,\rho_+^2\spdown)$ &  $(\rho_+\spup,-\rho_-\spdown)$ \\ \hline			
    $\mathbf{D}_{6h}$\hspace*{2mm} & $\Gamma_7$ & 2 & $(\spup,\spdown)$ & $if_{\Gamma_1^-}(\br)(\spup,\spdown)$ \\ 
                      & $\Gamma_8$ & 2 & $(\rho_+^2\spup,\rho_-^2\spdown)$ & $(\rho_-^3\spup,-\rho_+^3\spdown)$ \\ 
                      & $\Gamma_9$ & 2 & $(\rho_-^2\spup,\rho_+^2\spdown)$ &  $(\rho_+\spup,-\rho_-\spdown)$ \\ \hline
    $\mathbf{T}_{h}$  & $\Gamma_5$ & 2 & $(\spup,\spdown)$ & $if_{\Gamma_1^-}(\br)(\spup,\spdown)$ \\
		      & $(\Gamma_6,\Gamma_7)$ & 4 & - & - \\ \hline                       
    $\mathbf{O}_{h}$  & $\Gamma_6$ & 2 & $(\spup,\spdown)$ & $if_{\Gamma_1^-}(\br)(\spup,\spdown)$ \\
		      & $\Gamma_7$ & 2 & $f_{\Gamma_2^+}(\br)(\spup,\spdown)$\hspace*{2mm} & $if_{\Gamma_2^-}(\br)(\spup,\spdown)$ \\ 
                      & $\Gamma_8$ & 4 & - & - \\ \botrule	
\end{tabular}}
\label{table: Gamma-point-coreps}
\end{table}

\section{Green's functions}
\label{app: GFs}

We introduce the normal and anomalous Green's functions in the Matsubara representation:\cite{AGD}
\begin{eqnarray}
\label{GFs-definitions}
  && G_{nn',ss'}(\bk,\tau)=-\langle T_\tau c_{\bk,n,s}(\tau)c^\dagger_{\bk,n',s'}(0)\rangle, \nonumber \\
  && F_{nn',ss'}(\bk,\tau)=\langle T_\tau c_{\bk,n,s}(\tau)\tilde c_{\bk,n',s'}(0)\rangle, \\
  && \bar F_{nn',ss'}(\bk,\tau)=\langle T_\tau \tilde c^\dagger_{\bk,n,s}(\tau)c^\dagger_{\bk,n',s'}(0)\rangle. \nonumber
\end{eqnarray}
These Green's functions form $2N\times 2N$ matrices, denoted below by $\check G$, \textit{etc}, in the direct product of the band and Kramers spaces. We reserve the notation $\hat G$ for $2\times 2$ matrices 
in the Kramers space. Introducing the four-component Nambu operators in the $n$th band,
$$
  C_{\bk,n}=\left(\begin{array}{c}
                  c_{\bk,n,1} \\
                  c_{\bk,n,2} \\
                  \tilde c^\dagger_{\bk,n,1} \\
                  \tilde c^\dagger_{\bk,n,2} \\
            \end{array}\right)
            =\left(\begin{array}{c}
                  c_{\bk,n,1} \\
                  c_{\bk,n,2} \\
                  p_nc^\dagger_{-\bk,n,2} \\
                  -p_nc^\dagger_{-\bk,n,1} \\
            \end{array}\right),      
$$
one can combine the expressions (\ref{GFs-definitions}) into
\begin{eqnarray}
\label{matrix G}
    {\cal G}_{nn'}(\bk,\tau) &=& -\langle T_\tau C_{\bk,n}(\tau)C^\dagger_{\bk,n'}(0)\rangle \nonumber\\
    &=& \left(\begin{array}{cc}
        \hat G_{nn'}(\bk,\tau) & -\hat F_{nn'}(\bk,\tau) \\
        -\hat{\bar F}_{nn'}(\bk,\tau) & \hat{\bar G}_{nn'}(\bk,\tau) \\
    \end{array}\right),
\end{eqnarray}
where the auxiliary normal Green's function is given by
\begin{eqnarray*}
  \bar G_{nn',ss'}(\bk,\tau) &=& -\langle T_\tau \tilde c^\dagger_{\bk,n,s}(\tau)\tilde c_{\bk,n',s'}(0)\rangle \\
  &=& -p_np_{n'}[\hat\sigma_2\hat G_{n'n}(-\bk,-\tau)\hat\sigma_2]_{s's},
\end{eqnarray*}
using Eq. (\ref{tilde-c}). 
 
The Green's functions satisfy the Gor'kov equations obtained from the Hamiltonian (\ref{H-general}) using the standard procedure:\cite{AGD}
\begin{eqnarray}
\label{Gorkov-Eqs}
  -\frac{\partial{\cal G}(\bk,\tau)}{\partial\tau} &=& \delta(\tau)\mathbb{1}_{4N} \nonumber\\
      && + \left(\begin{array}{cc}
               \check\epsilon(\bk) & \check\Delta(\bk) \\
               \check\Delta^\dagger(\bk) & -\check{\tilde\epsilon}(\bk)
        \end{array}\right){\cal G}(\bk,\tau),
\end{eqnarray}
where $\mathbb{1}_{4N}$ is the $4N\times 4N$ unit matrix, 
\begin{eqnarray*}
  && \epsilon_{nn',ss'}(\bk)=\xi_n(\bk)\delta_{nn'}\delta_{ss'}-\bH\bm{m}_{nn',ss'}(\bk),\\
  && \tilde\epsilon_{nn',ss'}(\bk)=\xi_n(\bk)\delta_{nn'}\delta_{ss'}+\bH\bm{m}_{nn',ss'}(\bk),
\end{eqnarray*}
and $\Delta_{nn',ss'}(\bk)=\Delta_{n,ss'}(\bk)\delta_{nn'}$.
Note that the gap functions are band-diagonal (since we neglected the interband pairing), whereas the single-particle energy $\check\epsilon$ and its time-reversed counterpart $\check{\tilde\epsilon}$ are not.
Using the Matsubara frequency representation, ${\cal G}(\bk,\tau)=T\sum_m{\cal G}(\bk,\omega_m)e^{-i\omega_m\tau}$, where $\omega_m=(2m+1)\pi T$, one can solve Eq. (\ref{Gorkov-Eqs}), with the following result:
\begin{eqnarray}
\label{GFs-k-omega}
  && \check G=[i\omega_m-\check\epsilon-\check\Delta(i\omega_m+\check{\tilde\epsilon})^{-1}\check\Delta^\dagger]^{-1}, \\
  && \check F=-(i\omega_m-\check\epsilon)^{-1}\check\Delta\check{\bar G}, \nonumber\\
  && \check{\bar F}=-(i\omega_m+\check{\tilde\epsilon})^{-1}\check\Delta^\dagger\check G, \nonumber\\
  && \check{\bar G}=[i\omega_m+\check{\tilde\epsilon}-\check\Delta^\dagger(i\omega_m-\check\epsilon)^{-1}\check\Delta]^{-1}. \nonumber
\end{eqnarray}
These expressions are valid at arbitrary magnetic field.

At zero field, $\check\epsilon$ and $\check{\tilde\epsilon}$ become band-diagonal. Therefore, the Green's functions are also band-diagonal: 
$G^{(0)}_{nn',ss'}=\delta_{nn'}G^{(0)}_{n,ss'}$, \textit{etc},
where
\begin{eqnarray*}
\label{GFs-0}
  \hat G^{(0)}_n &=&-(i\omega_m+\xi_n)(\omega_m^2+\xi_n^2+\hat\Delta_n\hat\Delta_n^\dagger)^{-1}, \\
  \hat F^{(0)}_n &=& \hat\Delta_n(\omega_m^2+\xi_n^2+\hat\Delta_n^\dagger\hat\Delta_n)^{-1} \\
    &=& (\omega_m^2+\xi_n^2+\hat\Delta_n\hat\Delta_n^\dagger)^{-1}\hat\Delta_n, \\
  \hat{\bar F}^{(0)}_n &=& \hat\Delta_n^\dagger(\omega_m^2+\xi_n^2+\hat\Delta_n\hat\Delta_n^\dagger)^{-1}=[\hat F^{(0)}_n]^\dagger. 
\end{eqnarray*}
In a weak field, one can expand Eq. (\ref{GFs-k-omega}), with the following result: 
$$
  \check G=\check G^{(0)}-\check G^{(0)}(\bH\check{\bm m})\check G^{(0)}-\check F^{(0)}(\bH\check{\bm m})\check{\bar F}^{(0)}+{\cal O}(H^2).
$$
Substituting this into Eq. (\ref{M-from-G}) and dropping the superscript ``$(0)$'', we arrive at the expressions (\ref{M_0}) and (\ref{chi-general}) for the spontaneous magnetization and the spin susceptibility, respectively.

\section{Interband susceptibility}
\label{app: interband chi}

We consider just one pair of bands ($n,n'=1,2$) with $\xi_2(\bk)>\xi_1(\bk)$ and focus on the terms with $n\neq n'$ in Eq. (\ref{chi-general}). Using Eq. (\ref{m_12}), 
we obtain the following expression for the interband contribution to the spin susceptibility:
\begin{eqnarray}
\label{tilde-chi-general}
  \tilde\chi_{ij} = 4T\sum_m\int\frac{d^3\bk}{(2\pi)^3}\bigl[P_{ij}(\omega_m^2-\xi_1\xi_2)-Q_{ij}\bigr] \nonumber\\
  \times\frac{1}{(\omega_m^2+\xi_1^2+|\Delta_1|^2)(\omega_m^2+\xi_2^2+|\Delta_2|^2)},
\end{eqnarray}
where $P_{ij}=a_ia_j+\bmb_i\bmb_j$,
$$
  Q_{ij}=(a_ia_j+\bmb_i\bmb_j)\re(\psi_1^*\psi_2)
$$
in the singlet case, and
\begin{eqnarray*}
  Q_{ij} &=& (a_ia_j-\bmb_i\bmb_j)\re(\bmd_1^*\bmd_2)\\
  && -(a_i\bmb_j+a_j\bmb_i)\re(\bmd_1^*\times\bmd_2)\\
  && +\re(\bmb_i\bmd_1^*)(\bmb_j\bmd_2)+\re(\bmb_i\bmd_2^*)(\bmb_j\bmd_1)
\end{eqnarray*}
in the unitary triplet case. In the normal state, the interband susceptibility (\ref{tilde-chi-general}) takes the form (\ref{chi-inter-normal}).
To calculate the interband susceptibility in the superconducting state, we assume the weak-coupling picture, so that the gap functions are nonzero only inside the BCS shells near the Fermi surfaces: 
$\psi_n(\bk),\bmd_n(\bk)\propto\theta(\epsilon_c-|\xi_n(\bk)|)$. For  a large band splitting, ${\cal E}(\bk)\gg\epsilon_c$,
the BCS shells in the bands 1 and 2 do not overlap and therefore $Q_{ij}=0$ in both singlet and triplet cases. 

One can expect that the interband susceptibility is almost unchanged when the system undergoes a superconducting phase transition, in which only the electrons near the Fermi surfaces are affected. 
To find the upper bound on the effect of superconductivity on $\tilde\chi_{ij}$, it is sufficient to calculate the latter at $T=0$, where the Matsubara sum in Eq. (\ref{tilde-chi-general}) can be replaced by a
frequency integral. In this way, we obtain:
\begin{eqnarray*}
  && \tilde\chi_{ij}^S(T=0)-\tilde\chi_{ij}^N=2\int\frac{d^3\bk}{(2\pi)^3}P_{ij} \\
  && \quad\times\biggl[\left(1-\frac{\xi_1\xi_2}{E_1E_2}\right)\frac{1}{E_1+E_2}-\frac{1-\sign(\xi_1\xi_2)}{|\xi_1|+|\xi_2|}\biggr],
\end{eqnarray*}
where the excitation energies $E_{1,2}$ are given by Eq. (\ref{excitation-energy}). 
The expression in the square brackets is nonzero only near the two Fermi surfaces, therefore the susceptibility deviation can be written as a sum of the independent contributions from the two BCS shells:  
$\tilde\chi_{ij}^S-\tilde\chi_{ij}^N={\cal I}_{ij}^{(1)}+{\cal I}_{ij}^{(2)}$, where
\begin{eqnarray*}
  && {\cal I}_{ij}^{(1)}=2\int\frac{d^3\bk}{(2\pi)^3}\theta(\epsilon_c-|\xi_1|)P_{ij} \\
  && \quad\times\left[\left(1-\frac{\xi_1}{E_1}\sign\,\xi_2\right)\frac{1}{E_1+|\xi_2|}-\frac{1-\sign(\xi_1\xi_2)}{|\xi_1|+|\xi_2|}\right]
\end{eqnarray*}
and ${\cal I}_{ij}^{(2)}$ is obtained from ${\cal I}_{ij}^{(1)}$ by replacing $\xi_1\leftrightarrow\xi_2$, $E_1\leftrightarrow E_2$. Note that $\xi_2>0$ in the BCS shell in band $1$, while
$\xi_1<0$ in the BCS shell in band $2$. Next, we neglect the energy dependence of the single-particle DoS, of $P_{ij}$, and of ${\cal E}$ near 
the Fermi surface and obtain: 
\begin{equation}
\label{cal-I_1}
  {\cal I}_{ij}^{(1)}=2N_{F,1}\langle P_{ij} I_1\rangle_{FS,1},
\end{equation}
where
$$
  I_1=\int_{-x_m}^{x_m} dx\left(\frac{1-x/\sqrt{x^2+\delta^2}}{1+x+\sqrt{x^2+\delta^2}}-\frac{1-\sign\,x}{1+x+|x|}\right),
$$
$x_m=\epsilon_c/{\cal E}$, and $\delta=|\Delta_1|/{\cal E}$. Since $\delta\ll x_m\ll 1$, the last integral can be easily calculated with the logarithmic accuracy: $I_1\simeq -2\delta^2\ln(x_m/\delta)$.
Substituting this into Eq. (\ref{cal-I_1}) and repeating the calculation for the second band, we finally obtain: $\tilde\chi_{ij}^S(T=0)-\tilde\chi_{ij}^N\propto(|\Delta|^2/{\cal E}^2)\ln(\epsilon_c/|\Delta|)$.

\section{Derivation of Eq. (\ref{chi-intra-triplet})}
\label{app: triplet chi}

Substituting the Green's functions (\ref{GFs-zero-field}) in Eq. (\ref{chi-intra}), the intraband susceptibility in the triplet state can be represented in the form
$\chi_{ij}=\chi^{(1)}_{ij}+\chi^{(2)}_{ij}$, where
$$
  \chi^{(1)}_{ij}=2T\sum_m\int\frac{d^3\bk}{(2\pi)^3}\frac{\omega_m^2-\xi^2-|\bmd|^2}{(\omega_m^2+\xi^2+|\bmd|^2)^2}\bigl(\hat\mu\hat\mu^\dagger\bigr)_{ij}
$$
and
$$
  \chi^{(2)}_{ij}=4T\sum_m\int\frac{d^3\bk}{(2\pi)^3}\frac{|\bmd|^2}{(\omega_m^2+\xi^2+|\bmd|^2)^2}\bigl(\hat\mu\hat \Pi_\perp\hat\mu^\dagger\bigr)_{ij},
$$
where $\Pi_{\perp,\mu\nu}=\delta_{\mu\nu}-\re(d_\mu d_\nu^*)/|\bmd|^2$. Summing over the Matsubara frequencies in $\chi^{(1)}$, we obtain:
\begin{equation}
\label{chi-triplet-1}
  \chi^{(1)}_{ij}=2N_F\bigl\langle\bigl(\hat\mu\hat\mu^\dagger\bigr)_{ij}{\cal Y}\bigr\rangle_{FS},
\end{equation}
where ${\cal Y}$ is the Yosida function (\ref{Yosida-function}) with $|\Delta(\bk)|=|\bmd(\bk)|$.

To express $\chi^{(2)}$ in terms of the Yosida function, we use the fact that $\bmd(\bk)$ is nonzero only inside the BCS shell near the Fermi surface, which allows one to write
$$
  \chi^{(2)}_{ij}=2N_F\langle (\hat\mu\hat \Pi_\perp\hat\mu^\dagger)_{ij}(J_+-J_-)\rangle_{FS}, 
$$  
where 
$$
  J_\pm=T\sum_m\int_{-\infty}^\infty d\xi\frac{\omega_m^2-\xi^2\pm|\bmd|^2}{(\omega_m^2+\xi^2+|\bmd|^2)^2}.
$$
It is easy to see that $J_+=1$ (this can be shown by adding and subtracting the value of $J_+$ in the normal state and performing the energy integration before the Matsubara summation)) and $J_-={\cal Y}$, 
therefore
\begin{equation}
\label{chi-triplet-2}
  \chi^{(2)}_{ij}=2N_F\bigl\langle\bigl(\hat\mu\hat \Pi_\perp\hat\mu^\dagger\bigr)_{ij}(1-{\cal Y})\bigr\rangle_{FS}.
\end{equation}
Combining this with Eq. (\ref{chi-triplet-1}), we obtain:
\begin{eqnarray}
\label{chi-intra-triplet-appendix}
  \chi_{ij}(T) = 2N_F\left\langle\bigl[\hat\mu(\bk)\hat\mu^\dagger(\bk)\bigr]_{ij}{\cal Y}(\bk,T)\right\rangle_{FS}\nonumber\\
  +2N_F\left\langle\bigl[\hat\mu(\bk)\hat \Pi_\perp(\bk)\hat\mu^\dagger(\bk)\bigr]_{ij}\bigl[1-{\cal Y}(\bk,T)\bigr]\right\rangle_{FS},
\end{eqnarray}
which becomes Eq. (\ref{chi-intra-triplet}) after rearranging the terms.


\begin{thebibliography}{99}

\bibitem{SU-review}
M. Sigrist and K. Ueda, Rev. Mod. Phys. \textbf{63}, 239 (1991).

\bibitem{TheBook}
V. P. Mineev and K. V. Samokhin, \textit{Introduction to Unconventional Superconductivity} (Gordon and Breach, London, 1999).

\bibitem{Kittel-book}
C. Kittel, \textit{Quantum Theory of Solids} (Wiley, 1987).

\bibitem{CB60}
M. H. Cohen and E. I. Blount, Phil. Mag. \textbf{5}, 115 (1960).

\bibitem{Roth66}
L. M. Roth, Phys. Rev. \textbf{145}, 434 (1966).

\bibitem{LL-9}
E. M. Lifshitz and L. P. Pitaevskii, {\em Statistical Physics, Part 2}, Sec. 56 (Butterworth-Heinemann, Oxford, 2002).

\bibitem{Winkler-book}
R. Winkler, \textit{Spin-orbit Coupling Effects in Two-Dimensional Electron and Hole Systems} (Springer, Berlin, 2003).

\bibitem{WW09}
Y. Wan and Q.-H. Wang, Europhys. Lett. \textbf{85}, 57007 (2009).

\bibitem{Fis13}
M. H. Fischer, New J. Phys. \textbf{15}, 073006 (2013).

\bibitem{RS16}
A. Ramires and M. Sigrist, Phys. Rev. B \textbf{94}, 104501 (2016).

\bibitem{NHI16}
T. Nomoto, K. Hattori, and H. Ikeda, Phys. Rev. B \textbf{94}, 174513 (2016).

\bibitem{j32}
H. Kim, K. Wang, Y. Nakajima, R. Hu, S. Ziemak, P. Syers, L. Wang, H. Hodovanets, J. D. Denlinger, P. M. R. Brydon, D. F. Agterberg, M. A. Tanatar, R. Prozorov, and J. Paglione, Sci. Adv. \textbf{4}, eaao4513 (2018).

\bibitem{Sam19}
K. V. Samokhin, Ann. Phys. (N. Y.) \textbf{407}, 179 (2019).

\bibitem{Sam19-PRB}
K. V. Samokhin, Phys. Rev. B \textbf{100}, 054501 (2019).

\bibitem{And84}
P. W. Anderson, Phys. Rev. B \textbf{30}, 4000 (1984).

\bibitem{UR85}
K. Ueda and T. M. Rice, Phys. Rev. B \textbf{31}, 7114 (1985).

\bibitem{NCSC-book}
\textit{Non-centrosymmetric Superconductors: Introduction and Overview}, ed. by E. Bauer and M. Sigrist, Lecture Notes in Physics \textbf{847} (Springer, Heidelberg, 2012). 

\bibitem{LL-3}
L. D. Landau and E. M. Lifshitz, \textit{Quantum Mechanics} (Butterworth-Heinemann, Oxford, 2002).

\bibitem{Bir-Pikus-book}
G. L. Bir and G. E. Pikus, \textit{Symmetry and Strain-induced Effects in Semiconductors}, Sec. 25 (Wiley, New York, 1974).

\bibitem{And59}
P. W. Anderson, J. Phys. Chem. Solids \textbf{11}, 26 (1959).

\bibitem{Blount85}
E. I. Blount, Phys. Rev. B \textbf{32}, 2935 (1985).

\bibitem{Leggett75}
A. J. Leggett, Rev. Mod. Phys. \textbf{47}, 331 (1975).

\bibitem{Lax-book}
M. Lax, \textit{Symmetry Principles in Solid State and Molecular Physics} (Dover Publications, New York, 2001).

\bibitem{UPt3-review}
R. Joynt and L. Taillefer, Rev. Mod. Phys. \textbf{74}, 235 (2002).

\bibitem{Strand09}
J. D. Strand, D. J. Van Harlingen, J. B. Kycia, and W. P. Halperin, Phys. Rev. Lett. \textbf{103}, 197002 (2009).

\bibitem{Schemm14}
E. R. Schemm, W. J. Gannon, C. M. Wishne, W. P. Halperin, and A. Kapitulnik, Science \textbf{345}, 190 (2014).

\bibitem{Avers20}
K. E. Avers, W. J. Gannon, S. J. Kuhn, W. P. Halperin, J. A. Sauls, L. DeBeer-Schmitt, C. D. Dewhurst, J. Gavilano, G. Nagy, U. Gasser, and M. R. Eskildsen, Nat. Phys. \textbf{16}, 531 (2020). 

\bibitem{Gannon17}
W. J. Gannon, W. P. Halperin, M. R. Eskildsen, P. Dai, U. B. Hansen, K. Lefmann, and A. Stunault, Phys. Rev. B \textbf{96}, 041111(R) (2017).

\bibitem{AG62}
A. A. Abrikosov and L. P. Gor'kov, Zh. Eksp. Teor. Fiz. \textbf{42}, 1088 (1962) [Sov. Phys. JETP \textbf{15}, 752 (1962)].

\bibitem{BD68}
C. J. Bradley and B. L. Davies, Rev. Mod. Phys. \textbf{40}, 359 (1968).

\bibitem{BC-book}
C. J. Bradley and A. P. Cracknell, \textit{The Mathematical Theory of Symmetry in Solids} (Oxford University Press, Oxford, 2010).

\bibitem{Lutt56}
J. M. Luttinger, Phys. Rev. \textbf{102}, 1030 (1956).

\bibitem{AGD}
A. A. Abrikosov, L. P. Gorkov, and I. E. Dzyaloshinski, \textit{Methods of Quantum Field Theory in Statistical Physics} (Dover Publications, New York, 1975).


\end{thebibliography}
\end{document}